\documentclass{eceasst}

\usepackage[colorinlistoftodos,prependcaption,textsize=tiny]{todonotes}
\usepackage{graphicx}
\usepackage{multirow}
\usepackage{underscore}
\usepackage{subfig}

\title{Who Do You Think You Are? Creating RSE Personas \\ from GitHub Interactions} 

\short{RSE Personas} 

\author{
Felicity Anderson\autref{1},
Julien Sindt\autref{1} and
Neil Chue Hong\autref{1}
}
\institute{
\autlabel{1} \email{felicity.anderson@ed.ac.uk}, \\
EPCC, University of Edinburgh, Edinburgh, UK\par}

\abstract{    
    We describe data-driven RSE personas: an approach combining software repository mining and data-driven personas applied to research software (RS), an attempt to describe and identify common and rare patterns of Research Software Engineering (RSE) development.
    This allows individuals and RS project teams to understand their contributions, impact and repository dynamics - an important foundation for improving RSE. 
    We evaluate the method on different patterns of collaborative interaction behaviours by contributors to mid-sized public RS repositories (those with 10-300 committers) on GitHub.
    We demonstrate how the RSE personas method successfully characterises a sample of 115,174 repository contributors across 1,284 RS repositories on GitHub, sampled from 42,284 candidate software repository records queried from Zenodo. 
    We identify, name and summarise seven distinct personas from low to high interactivity: Ephemeral Contributor; Occasional Contributor; Project Organiser; Moderate Contributor; Low-Process Closer; Low-Coding Closer; and Active Contributor. 
    This demonstrates that large datasets can be analysed despite difficulties of comparing software projects with different project management factors, research domains and contributor backgrounds.
    } 


\begin{document}
\maketitle

\section{Introduction}\label{Intro}

    There are many benefits to understanding different types of contributors to the collaborative development of Research Software (RS).
    The ability to describe individual and team impact is important for receiving credit (both social and professional), planning and allocating work, and identifying potential skill gaps in teams.
    However, estimation of coding contributions in collaborative software projects is a known difficulty; much effort within software engineering research has been dedicated to the generation of metrics, methods, and models to quantify this (\cite{miranda_quantitative_2025,de_bassi_measuring_2018,lima_assessing_2015}).
    This study proposes a novel approach combining two data-focused methods from traditional Software Engineering (SE) research and applying them to the field of Research Software (RS), in the attempt to characterise Research Software Engineers (RSEs) contribution behaviours on RS projects using newly developed \textit{RSE Personas}. 
    
    Personas are used in traditional SE design to describe groups of users with similar properties during product design.
    The Persona methodology has been further developed to describe groups and patterns derived from complex datasets. 
    These Personas might be used for example to describe groups of users with similar properties, priorities and needs; helping to summarise the group's characteristics.

    This study focuses on RS and those who contribute to its' creation, adopting the definition of \textit{Research Software Engineers (RSEs)} from \cite{community_turing_2025}: ``Research Software Engineers (RSEs) are programmers with scientific backgrounds who play increasingly critical roles in the conduct of research and production of research software tools.'' 
    This research also recognises non-coding contributors (i.e. researchers and project managers) to RS projects under the term RSE: their engagement with the project helps shape the software developed, and therefore are critically important, though potentially harder to quantify from RS repository data.
    The RSE community is a broad one, and consists of members from a wide variety of professional and academic backgrounds; the term `RSE' was settled upon in 2012 to encompass the fusion of these worlds in one role (\cite{hettrick_not-so-brief_2016}). 
    RSE Personas follow on from the idea that utilising descriptive names for useful concepts allows people contributing to RS to adopt and discuss them, pushing the RSE community further.
    The diversity of skills and domains encompassed by RSEs could result in a varied array of ways of working within their RS projects - this study explores whether different patterns of interaction behaviour exist, and just how diverse they might be. 

    Mining RS repositories provides a rich dataset of interactions evidence to explore how RSEs engage with their projects. Zenodo\cite{european_organization_for_nuclear_research_zenodo_2013} is an open research and data repository which holds hundreds of thousands of research software records, and grants a digital object identifier (DOI) for each deposited object. 
    These DOIs enable links to related datasets, code and publications.
    As such, Zenodo represents an effective tool for sourcing RS GitHub (GH) repositories for software repository mining. 
    When combined with GitHub's API and large source of repository data and incorporated interactions data, these two platforms support a powerful workflow for creating Data-Driven Personas generated from collaborative RS repository data.
    
    GitHub is the most commonly used version control platform amongst researchers, although research institutes may host their own instances of other platforms (e.g. GitLab, BitBucket) for private code. 
    These collaborative platforms often offer additional benefits including code- and project management- tools such as issue tickets, code review through pull requests, and release tagging.
    Many RS projects have adopted these functions, particularly amongst larger teams with RSEs, where file sharing and change tracking are crucial for effective dispersed working.
    While GH offers many benefits to its' users, it also enables powerful research into the footprints left by contributors' engagement with repositories in the form of commits or issue tickets, for example.
    These interactions supply insights into who does what, and how: the ideal empirical basis for data-driven methods for characterising patterns and groupings such as Personas.

    In this paper, we introduce \textbf{\textit{data-driven RSE Personas}} -- a novel application of the Personas method to research software -- and explore how they can be generated from RS repository data to explore development behaviour and how contributors engage with their project repositories. 

    The seven identified RSE Personas, from three initial Interactivity level groupings (Low, Moderate and High) are described and summarised in Section~\ref{PersonasSummary}. 
    \textit{Low Interactivity Personas} are: \textbf{Ephemeral Contributor} and \textbf{Occasional Contributor}, differentiated by their length of engagement with the repository, and the volume of their interactions: extremely low versus low (Ephemeral and Occasional, respectively).
    \textit{Moderate Interactivity} is demonstrated by the \textbf{Project Organiser} and \textbf{Moderate Contributor} personas, but these are separated by a preference for development management features over coding in Project Organiser, versus moderate coding enabling good Pull Request and Issue Ticket closure rates amongst Moderate Contributors. 
    \textit{High Interactivity} can be demonstrated in each of the three final personas: \textbf{Low-Process Closer} (showing higher programming-focused activities) and \textbf{Low-Coding Closer} (showing more development management feature use) show contrasting approaches favouring different interaction types, while \textbf{Active Contributor} demonstrates the overall highest repository interactions in terms of volume and variety of all RSE Personas.

    We discuss similar studies and important foundational literature in \nameref{RelatedWork} to support our aims and selected approaches, however more detail on these are given in \nameref{methods}, describing the software repository mining method used to collect the data, our key interaction variables, and the persona creation methods and their evaluation. 
    \nameref{ResultsAnalysis} presents patterns of contributions identified from the interactions data, and the seven RSE Personas which represent these patterns of behaviour.
    \nameref{Discussions} explores potential interpretations and limitations of the study. 
    In \nameref{FutureWork} we consider work which could follow on from this study and how these results may be used, while we restate our summarised findings in \nameref{Conclusion}.
    The code\footnote{Code in open repository on GitHub: \url{https://github.com/FlicAnderson/RSE-personas}.} and data\footnote{Data deposited on Zenodo: \url{https://zenodo.org/records/15458472}} from this research are open and publicly available; more details on the dataset, computational set-up and codebase are held in an \nameref{code-data}.
    All Figures are located after the Appendix, while Tables are throughout the text.

\section{Related Work}\label{RelatedWork}
    
    \cite{vidoni_systematic_2022} is a systematic literature review into Mining Software Repositories (MSR), providing guidelines for planning and carrying out MSR studies in ways which will avoid or mitigate potential `perils' noted by previous authors such as \cite{kalliamvakou_promises_2014}.
    These latter authors focus on using GitHub for MSR studies, and how SE researchers might obtain the benefits of large scale data collection but avoid the possible pitfalls of complex data, problems posed by techniques such as rebasing or commit squashing, or analysing forks and base repositories.
    The research by \cite{alrashedy_how_2024} into GitHub repositories which accompanied published research (which our study would classify as RS) looked to address questions relating to GH repository usage for publication-supporting-code over time, rates of interactions and responsiveness between the authors of such research and the public, and whether publishing code in this way contributed towards publication impact. 
    Their study found that only in a few cases did publication of code on a collaborative platform such as GH lead to external contribution by those who discovered the research; while those publishing their code did not respond to contributions to their repositories swiftly, or indeed at all, in many cases. 
    While those authors did investigate interaction types (issue ticket creation and closure, and pull request creation and closure), they note only that typically results indicate issue ticket and pull request interactions were both common (with issue tickets more frequent than pull requests), and repositories tended to receive either a feast or a famine - that is, ``either a few or hundreds such interactions''.
    Interestingly, the response by the publishing authors to engagement with their repository switched these two categories: repository owners more frequently responded to pull requests (89.9\%) than issues (64.8\%) from external contributors.
    
    \cite{milewicz_characterizing_2019} also used MSR methods, applied to 1863 open source scientific software projects obtained from the Journal of Open Source Software (JOSS), DOECODE (a public research platform), and GitHub (narrowed by searching for topics). 
    Examining a subset of 5000 contributors from across those population repositories, they explored contribution patterns by differing levels of researcher roles (combined with qualitative survey data of 72 developers of repositories from a subset of the study population).
    \cite{milewicz_characterizing_2019} found that senior research staff contributed an average 72\% of commits within their repositories and engaged with more technical file types in their commits. 
    Contribution periods were analysed using Kaplan-Meier survival analysis, defined as how long it takes a contributor to become inactive - in this case if a commit has not been made within the last 180 days via \cite{lin_developer_2017}. 
    This research indicates that different RS project roles appear to dictate contribution volumes and types of interactions: an important finding which influenced our study design. 
    From this paper we adopt a similar MSR methodology (although restricting our RS projects only to GH for purposes of efficiency) and also adopt the commit classification methods (\cite{hattori_nature_2008,vasilescu_variation_2014}) used by those authors, including their modifications to tailor them to RS. 

    \cite{dworatzyk_decoding_2024} explored the German software development community through the Stack Overflow survey results, and used clustering analysis methods to characterise differences between key groupings (personas) which had been identified. 
    The properties of the sub-groups were analysed for relationships to variables around community membership and participation: representing another approach towards investigating development interactions. 
    The clustering methods from that paper influenced this study, particularly in the choice of clustering algorithm, and treatment of data before clustering.    
    
    The data-driven personas approach used in this paper follows the recommendations made to practitioners implementing personas in the study into constructing data-driven personas by \cite{salminen_survey_2021}, in that it applies clustering methods, works with majority and minority subsets separately, and aims to be conscious of the potential for algorithmic biases.
    That research strongly influenced the design decision to work with the initial clusters separately and re-apply clustering methods to drill down further into more detailed differences within these initial sets.

\section{Methods and Dataset}\label{methods}

    This section outlines our \nameref{RQs-and-Hyps} and hypotheses in this research.
    We detail the \nameref{subsubsect:datacollection}, \nameref{subsubsect:sampling} and \nameref{subsubsect:analysismethods} approaches we have used to explore our variables and concepts of interest, including our method for generating and evaluating RSE Personas. 
    \autoref{fig:data-collection-workflow} describes this workflow in overview.     

    \subsection{Research Objectives and Research Questions}\label{RQs-and-Hyps}
    The objectives of this study are to investigate to what extent a data-driven Personas approach can be applied within the context of collaborative open (public) RS development on GitHub.

    The objectives motivated the following research questions: 
    
    \begin{itemize}
      \item RQ1: Can separate RSE Personas be identified and described from developer/repository interaction data?  
      \item RQ2: Which factors of RS repository interaction data most strongly influence the creation of RSE Personas?  
    \end{itemize}

    This research initially expected to find four main RSE Personas, each relating to a different combination of high and low values within two variables: the \textbf{Mean Repository Contributions (MRC)} is a general mean of all interaction types and \textbf{Unique Interaction Types (UIT)} are the sum of different types of interaction made by contributor. 
    These RSE Personas were hypothesised to be: 
    `Active Leaders' (high MRC and high UIT); `Project Managers' (low MRC and high UIT); `Occasional Contributors' (low MRC and low UIT); and `Focused Developers' (high MRC and low UIT). 

    These hypotheses were investigated within a pilot study carried out and presented in a poster for deRSE2025 \cite{anderson_who_2025}. 
    That study found only two of these Personas: `Active Leaders' and `Occasional Contributors'.
    Based on feedback received at deRSE2025 and further work with a considerably larger sample dataset, these hypotheses were updated for this study to determine whether further detail could be gathered.
    
    The hypotheses of this research were: 
    \begin{itemize}
        \item H1: At least two main RSE Personas can be identified: high interactivity and low interactivity, in line with the findings of the pilot study. 
        \item H2: High-responsibility type interactions (such as issue assignment, pull request closure) strongly contribute to RSE Persona separation.   
        \item H3: Further clustering, particularly within the low interactivity Persona, indicates the existence of additional Personas.
    \end{itemize}     
    
  \subsection{Study Methods}

    \subsubsection{Data Collection}\label{subsubsect:datacollection} 
    
    \paragraph{Evidence Required for Study Objectives} 
     
    The goal of our data collection was to obtain data relating to the types and scale of interactions made by contributors, during the process of developing RS in GH repositories. 
    Ensuring that selected repositories fulfilled the requirements of being open and accessible, and contained research software, was critical for research validity.  
    Repositories needed to primarily contain engineered RS code artifacts, rather than data, manuscripts, or other research software related artefacts, as discussed in \cite{munaiah_curating_2017}. 
    This was considered at the sample repository inclusion/exclusion stage (see Section \ref{criteria}). 

    The aims of this paper are supported by the following sampling strategy, using reasonable attempts to mitigate bias. 
    Recommendations for developing and reporting sampling strategies were followed in this study \cite{baltes_sampling_2021}, and we used a multi-stage sampling algorithm to obtain our sample.
    All data collection and processing code for this study can be found in the study GitHub repository (see Section \ref{code-data}) and the broad steps are described in Figure~\ref{fig:data-collection-workflow}.
   
    \paragraph{Gathering Candidate RS Repositories from Zenodo}
    
    The Zenodo REST API was used to source 42,284 candidate GH RS repositories from the Zenodo research repository in mid March 2025. 
    A query was submitted to the API to gather records of the type `software' with `open' accessibility with a related URL containing `github.com', returning 42,284 Zenodo software records matching the query.
    
    Not all records regarding GH repositories obtained through the Zenodo API were usable; from the 42,284 Zenodo IDs returned from the initial querying, only 39,288 records (92.91\%) returned full metadata information upon querying the individual record ID.

    \subsubsection{Data Sampling}\label{subsubsect:sampling}

    \paragraph{Mining 1284 RS Repositories for Developer Interactions Data}
   
    The GH API was used to confirm that the 39,288 GH repository names obtained via Zenodo a) remained on GH, and, b) were publicly visible, leaving a potential repository population of 24,955 GH repositories.
    
    Each of these accessible RS repositories was further queried against the study's inclusion/exclusion criteria (see Subsection \ref{criteria}), which enabled the final research dataset to be generated from eligible RS repositories which met those criteria.
    
    Only 2,981 (8.23\%) of these 24,955 repositories were eligible for inclusion in the study. 
    These were sub-sampled to generate the study sample of 1,284 repositories (a random selection of 45\% of these repositories) and were used to form the study dataset (the data and properties associated with those 1,284 included RS repositories).
    The 45\% proportion was selected as this was the maximum dataset size which could be successfully clustered with the methods described in Subsection~\ref{subsubsect:analysismethods} and our Computational Environment (described further in Appendix~\ref{code-data}).

    \paragraph{Inclusion/Exclusion Criteria}\label{criteria}

    These GitHub repositories were passed through an inclusion/exclusion step based on querying summary repository information against requirements to obtain the study sample of 2,981 RS GH repositories which fulfilled the following criteria: 

    \begin{enumerate}
        \item \textbf{Repositories must be publicly visible.}
        \item \textbf{The number of unique committers to repository must be at least 10 to ensure that they demonstrate the collaborative behaviour under study.}
        \item \textbf{The number of unique committers to repository must be less than 307 (2 standard deviations higher than the mean repository size).}
        \item \textbf{Repositories must have a permissive license, visible through the GH API.}
        \item \textbf{Repository languages must include at least one language popular within RS, as listed in \cite{alnoamany_towards_2018}'s list of the top 10 programming languages in a sample of RS.}
        \item \textbf{Repositories must not be a fork of another repository as these may indicate different interaction patterns.}
        \item \textbf{Repositories must be over 1000 days old to avoid `newly created' repositories which may show differences from established collaborative projects.}
    \end{enumerate}

    The mean age of population repositories was 1,865 days old. 
    The 'valid sample' list of 2,981 repositories who met all of these criteria were then randomly sampled (using a set seed) to obtain a percentage (45\%) of the total sample which was used for this analysis. 
    This sub-sampling was done for convenience as the clustering analysis stages are computationally intensive.
    
    The dataset used in this study therefore consists of 1,284 RS repositories (and 115,174 individuals).
    General summary information and interactions data were collated and analysed for all 115,174 repo-individuals who have contributed to these 1,284 sample repositories; comprising GH users who created issue tickets, commits information, and pull requests. 
    
    \paragraph{Key Variables}\label{KeyVars}
    
    As this study focuses on collaborative interactions within repositories, it was vital to understand and conceptualise three elements: the \textit{unit of study} (the contributor or RSE making the interactions); the relevant \textit{environment} (the RS GH repository with which interactions occurred); and the \textit{types of interactions} themselves.    

    MRC and RC of Pull Requests Created were found to be significant variables during the pilot study, explaining a high amount of variance in PCA analysis, and therefore seem to be a robust basis for RSE Personas creation.
    UIT, however, was found not to explain cluster variance well, possibly due to the differing `combinations' of possible interaction types. 
    Based on this, we analysed interaction types in more detail by looking at RC values separately and excluding UIT from clustering, though this concept remains useful in summarising variety of interaction types.
       
    \subparagraph{Repo-Individuals: the unit of study}

    This study separated `unique contributors' (unique GH usernames or GH user IDs) from `repo-individuals', defined here as a unique GH username + repository name combination (abbreviated from `repository-individual'). 
    This method enables the capture of changes in behaviour of GH users as they work across different repositories, each with different explicit project management approaches.

    \subparagraph{Included Interaction Types}
    The interaction types included in this study were: Commit Creation, Issue Ticket Creation, Issue Ticket Closure, Assignment to Issue Tickets, Pull Request Creation, Pull Request Closure.    
    Each of these interaction types were calculated as percentage Repository Contributions (RCs), and used in clustering methods (see Table \ref{tab:key-vars-concept}).

    \subparagraph{Interaction Days}
    Interaction days are defined as the sum of unique dates on which at least one included interaction type was contributed by a repo-individual. 
    As RSEs are suspected to have `gappy' contributions, calculating days on which they were active within the project gives a better general sense of intensity of interaction than the raw number of interactions alone or duration of involvement on a repository. 
    A repo-individual's share of interaction days over the repository lifetime can be calculated as Percentage Interaction Days.

        \begin{table}[]
    \centering
        \resizebox{\textwidth}{!}{%
        \begin{tabular}{ | p{0.25\linewidth} | p{0.6\linewidth} | }
        \hline
            \textbf{Variable} & \textbf{Description: ``The Percentage (\%) of..."} \\ \hline
            RC Commit Created & \ldots repository's commits created by repo-individual \\ \hline
            RC Issues Created & \ldots repository's issue tickets created by this repo-individual \\ \hline
            RC Issues Closed & \ldots repository's closed issue tickets which were closed by repo-individual \\ \hline
            RC Issues Assigned of Assigned & \ldots repository's assigned issue tickets which were assigned to this repo-individual \\ \hline
            RC Pull Request Created & \ldots repository's pull requests created by repo-individual \\ \hline
            RC Pull Request Closed & \ldots repository's pull requests closed by repo-individual \\ \hline
            MRC: Mean Repository Contribution & \ldots contribution across all repository interactions for this repo-individual. RC values for all interaction types are summed, and divided by the number of included interaction types to obtain a meta-average, showing the typical interactivity level of this repo-individual. If no contributions made for an interaction type zeroes are used; divisor is always 6 - the number of interaction types \\ \hline
            Percentage Created-Closed Issues & \ldots RC of Issues Created by this repo-individual minus RC of Issues Closed by them; indicates whether they are a net creator or closer of issues. \\ \hline    
            Percentage Sum N Interactions & \ldots repository's total interactions which were contributed by this repo-individual \\ \hline
            Percentage Interaction Days & \ldots the Sum of Interaction Days of all repo-individuals for this repository, indicating the proportion of time contributed by this repo-individual compared to other contributors. \\ \hline    
        \end{tabular}
        }
        \caption{Clustering variables: each variable represents a percentage related to each Repo-Individuals contributions to their repository, allowing contributors from different repositories to be compared. RC = ``Repository Contribution"}
        \label{tab:key-vars-concept}
    \end{table}
    
    \subparagraph{Interaction Variables Selection Details}
    
    Commit creation is a basic interaction with a GH repository and represents actual development - writing or editing code or other files.    
    
    Creating an issue ticket links a repo-individual to work around the codebase, but provides complimentary information about development compared to commits, instead relating to planning and adding new features, organising development tasks, identifying bugs or problems within the software, or tickets may be linked to more management-related aims. 
    Creating an issue ticket was also not assumed to be equivalent to working on that issue.

    Issue Ticket Closure is an important interaction type that relates more strongly to development management behaviours and could be considered a proxy (albeit a flawed one) for development effectiveness. 
    RSEs who close more issues could be presumed to have completed more coding tasks, or triaged them for relevance. 
    The net number of issues opened to issues closed indicates whether a repo-individual has a positive or negative balance of issue tickets to their name for a given project. 

    Issue tickets can each be assigned to one or more particular GH users by username. 
    Any member of an RS repository team can assign themselves or any other team member to an issue ticket, but no confirmation is required for tickets to be assigned. The person who is assigned may not have agreed to be assigned. 
    This study did not assume that assignment indicated acceptance of the task.
    
    A repo-individual assigned to an issue ticket is considered an interaction relating to the individual being assigned, and not the person assigning them. 
    Assigning an issue ticket indicates the assigner's opinion of the assignee: that this repo-individual is capable for this type of task, is available to carry out this work, or has the required authority or domain knowledge to accomplish it. 
    These assumptions provide valuable information about the assignee, but these are only assumptions by the assigner, and may not accurately reflect the true skill-set, availability or capabilities of the assignee. 

    Pull Request opening and closure is a similar yet stronger indicator of project responsibility to issue ticket data as not all project members have repository permissions allowing them to merge and close pull requests.

    \subsubsection{Data Analysis}\label{subsubsect:analysismethods}
    
    \paragraph{Interactions Analysis}\label{InteractionsAnalysis}

    Repo-individuals' `variety' of interactivity was examined to determine how spread or focused they were on types of interactions - contributing a high or low number of different types of interaction. 
    For example a repo-individual might contribute towards all six interaction types, showing a high breadth of interaction; or only generating issue tickets, showing narrow interactivity. 
    
    Interactivity could also be considered by `volume or depth' for a specific interaction type. 
    Contributing a high proportion of their repository's interactions would give high \%RC (Repository Contribution) values for that interaction type. 
    Likewise, fewer contributions of that type shows shallow interactivity and is indicated by a low \% RC.
    
    RC values for all interaction types were combined to indicate overall Mean Repository Contributions (MRC) by repo-individuals towards their repositories, and these MRC values describe the general trend of interactivity or `impact' made by a repo-individual. 
    
    Commit data was categorised according to two distinct methodologies, \cite{hattori_nature_2008}, and \cite{vasilescu_variation_2014}. 

    \subparagraph{Hattori-Lanza Commit Message Classification Method: Development Type}
    The \cite{hattori_nature_2008} method classifies commits according to initial keyword matching of commit message text against sets of verb stems.  
    This method generates a hierarchy of two main categories (\emph{Development} and \emph{Maintenance}) further split into more detailed classifications, and a remainder of unclassified commits which did not match any of the stem words (including commits with an empty message).
    
    \subparagraph{Vasilescu et al. Commit File-type Classification: Activity Types}
    The \cite{vasilescu_variation_2014} commit classification method instead uses regular expression style pattern matching to determine which `Activity Type' the commit should be given by matching the file-type of files changed in each commit against sets of file extensions associated with specific activities. 
    These pattern matching rules are applied in order, and classify each commit into one a list of Activity Types.
    
    \paragraph{Data-Driven Persona Creation}\label{PersonasMethods}

    \subparagraph{Selecting Hierarchical Clustering Method for Persona Generation}
    
    Hierarchical Clustering was used to develop RSE Personas, applying the \texttt{AgglomerativeClustering} function to pre-processed per-repo-individual data using the \texttt{sklearn.cluster} python package, using Euclidian distances and Ward algorithm. 
    All variables are already normalised as percentages, so no standardisation or encoding steps were required pre-clustering.
    
\begin{figure}[ht]
        \centering
        \subfloat[
        Workflow generating a study dataset of 115,174 repo-individuals across 1,284 RS repositories.
        \label{fig:data-collection-workflow}
        ]{
            \includegraphics[width=0.35\textwidth]{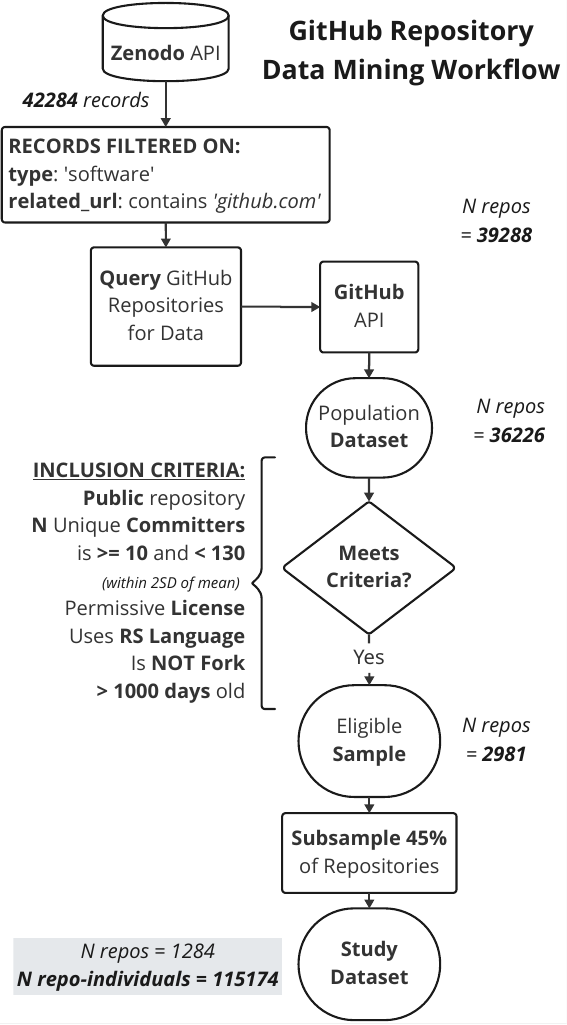}
        }
        \quad
        \subfloat[
        CH Index values indicated that data should be assigned to three initial clusters to best explain relationships in the data without over-fitting.\label{fig:CH-index}
        ]{
            \includegraphics[width=0.4\textwidth]{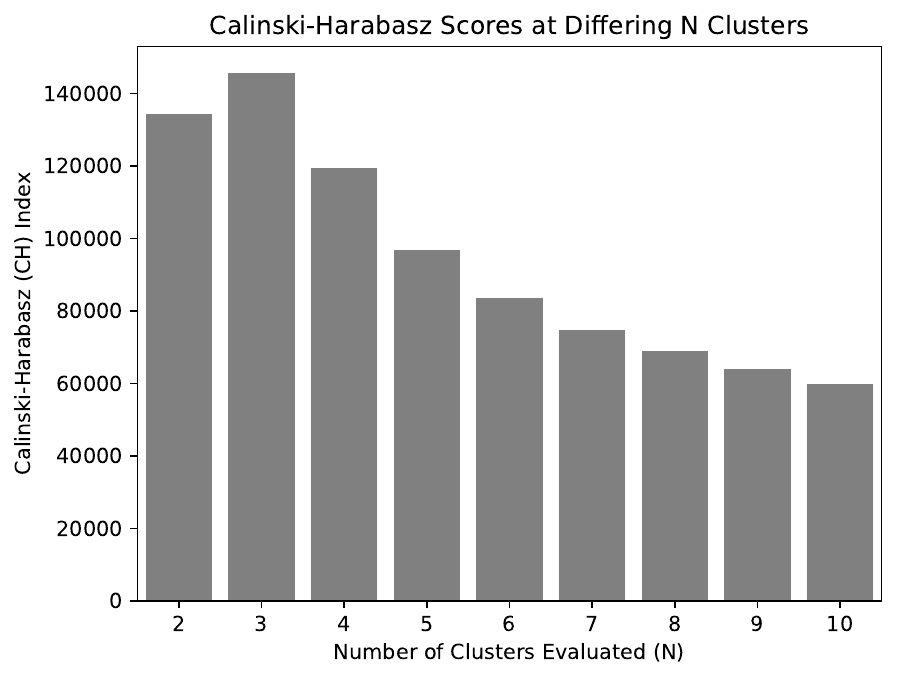}
        }
        \quad
        \subfloat[
        Dendrogram generated from 115,174 repo-individuals via hierarchical clustering analysis using Ward method and Euclidean distance, showing, from left to right: Cluster 1, then Cluster 2 and Cluster 0 are more closely related by the clustering variables. Vertical distance shows degree of differentiation between branches from the same node. Cluster 1 (found to feature Highest Interactivity) is most distinct with a long vertical parent branch, while Clusters 2 and 0 (Low and Moderate Interactivity respectively) are more similar to each other.\label{fig:dendro}
        ]{
            \includegraphics[width=1\textwidth]{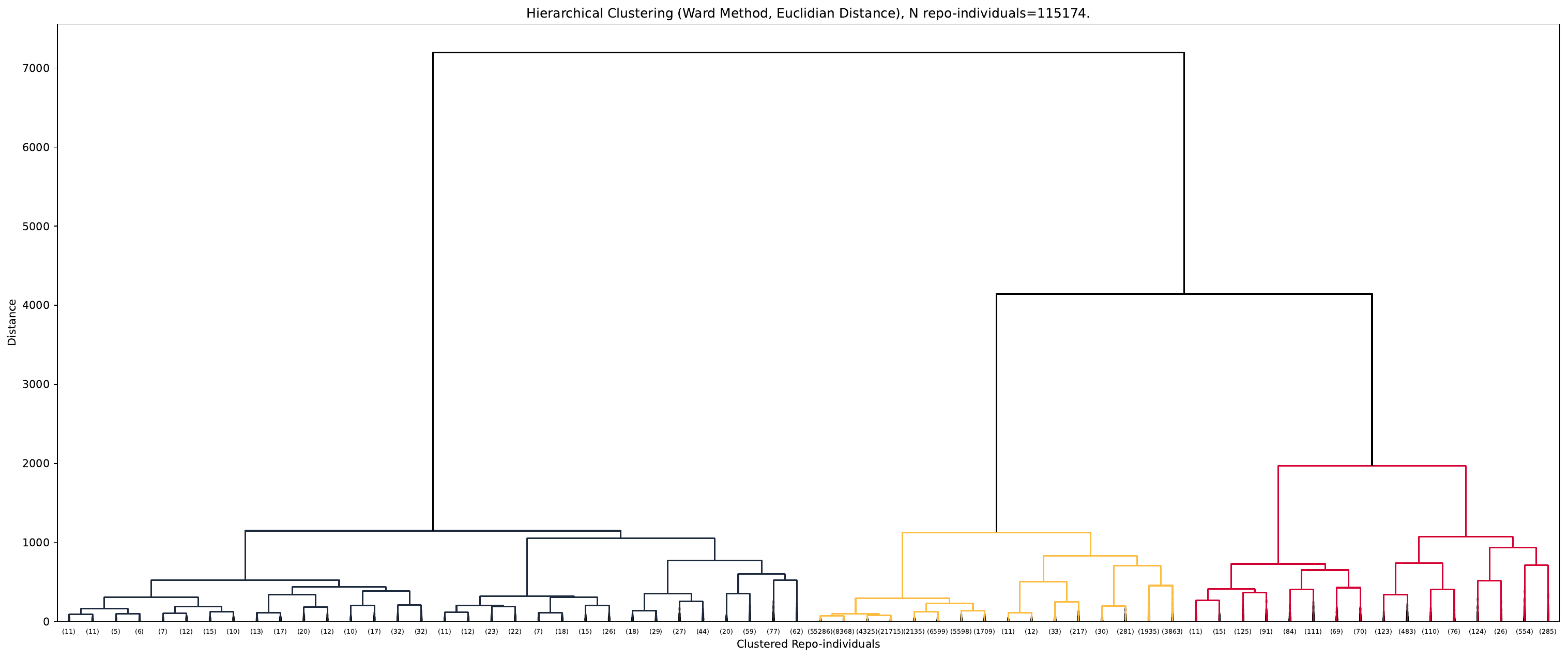}
        }
        \caption{Data collection and sampling workflow (\autoref{fig:data-collection-workflow}); Evaluation methods including calculating CH Index (\autoref{fig:CH-index}) and visualisation with a dendrogram (\autoref{fig:dendro}).}\label{fig:eval-methods-CH-dendro}
    \end{figure}

    The Calinski-Harabasz (CH) Index \cite{calinski_dendrite_1974} was used to evaluate the best number of clusters to describe groups of variation without over-fitting from the data by calculating the CH index for up to ten clusters. 
    N=3 clusters showed the highest CH score (CH index=145,660.97), so was used for the final analysis clustering - see Figure~\ref{fig:CH-index}.
    
    Additionally, a dendrogram (\autoref{fig:dendro}) was generated from the data and compared against the CH Index scores to confirm which number of clusters to generate for the initial clustering.
    
    \subparagraph{Minority Subsets: Working with Minority / Majority Personas Separately}  
    The three initial clusters identified from the dendrogram and generated using hierarchical clustering were further analysed to identify whether sub-clusters existed and address hypothesis H3.
    The proportions of the sample these clusters occupy can be seen in \autoref{fig:flowchart-intial-clusters} as a flowchart.
    Each initial cluster was subset out, re-clustered and analysed further (Figures \ref{fig:flowchart-cluster0} to \ref{fig:flowchart-cluster2}).

    \subparagraph{Validating Emerging Personas}

    Before clustering, the included interaction type RC scores were plotted to manually identify potential groupings which may represent clusters, iteratively feeding the decision into which variables to use in clustering. 
    
    Maximum \textbf{Calinski-Harabasz (CH) Index Scores} were calculated for numbers of sub-clusters to generate using hierarchical clustering: at N=2 sub-clusters (CH Index = 830.79) for Moderate Interactivity Cluster 0; N=3 for High Interactivity Cluster 1 (CH Index = 165.55), and N=2 for Low Interactivity Cluster 2 (CH Index = 39,192.63).
    
    \textbf{Principal Component Analysis (PCA)} was used to visualise resulting clusters by reducing dimensionality of the dataset to three eigenvectors summarising variance (see Figure~\ref{fig:pca-3d}), indicating the degree of separation, spread and overlap between potential clusters.
        
    \textbf{Feature Importance Analysis} (results in Table~\ref{tab:feature-importance}) was performed to identify which variables explained the greatest variance for each principal component (equivalent to the PCA axes).
    The feature importance for the two highest principal components were cross checked to confirm they showed the expected relationships.
    
    For variables with strong Feature Importance Analysis scores, \textbf{One-Way Analysis of Variance (ANOVA)} was used to confirm statistical significance using the \texttt{scipy.stats} module \texttt{f\_oneway}. 
    It should be noted that this module reports very small p values as 0.0 as they are hard to distinguish from zero.
        
    Tukey's Honest Significant Difference Test (Tukey's HSD) was used in a similar way to ANOVA but gave greater detail about the way in which clusters differed from each other and the significance of those relationships.

\section{Results and Analysis}\label{ResultsAnalysis}

    In \nameref{subsect:characterisingdata}, we explore the RS projects population and characteristics of their contributors before and after sub-sampling to create the analysis dataset.

    We outline initial clustering results and how we have analysed and evaluated these, exploring first the \nameref{subsect: VarietyInteractions}, then look at the scale of contributions in \nameref{subsect: VolumeInteractions} and how these dimensions help separate the patterns of interactions and contributions to generate our \nameref{PersonasSummary}. 
    There, we describe each of the \textit{seven identified RSE Persona} and their interrelationships from Low to High overall Interactivity: \textbf{Ephemeral Contributor}, \textbf{Occasional Contributor}, \textbf{Project Organiser}, \textbf{Moderate Contributor}, \textbf{Low-Process Closer}, \textbf{Low-Coding Closer}, \textbf{Active Contributor}.

    \subsection{Characterising the Data}\label{subsect:characterisingdata}

    \subsubsection{Characterising the RS Population Chosen for Sampling}\label{ThePopulation} 

    Collaborative RS projects are far from the norm on GH -- \textbf{the most common (mode) RS repository within the pre-sampled population has a single developer, three commits to the main branch, is 50 days old, and is currently inactive}.
    
    The majority (15,405) of the 36,226 RS repositories from GH have only one contributor (defined by GH API as the number of unique GH usernames which contributed at least 1 commit to the repository). 
    The remainder of the population demonstrate a swift drop into a long tail of increasing contributor numbers.
    4,911 repositories contained more than 10 contributors, while only 19 contained more than 1,000 developers.

    59.51\% of repositories have fewer than 100 commits on the main branch, while 20\% had less than 10 commits on the main branch, suggesting low activity is normal within the population. 
    The mean number of commits on the main branch across the population was shown to be 575.40.
    Meanwhile, just under half of population repositories (47.63\%) contained one or more commits made within the last year, with a mean of 36 commits in the last year per repository.
    
    Only 31.78\% of the population (11,514 repositories) had one or more open issue tickets, with the mean number of open tickets being 11 tickets.
    44\% of population repositories has one or more closed issue tickets, with a mean number of 107.72 closed tickets.    
    Only 17,125 repositories have pull requests enabled for their repositories. 


    \begin{figure}[!tbp]
      \centering
      \subfloat[Population repository sizes by log number of contributors; most are `lone-wolf' repositories with single contributors, the median is two, and the population mean is just over nine (9.32) contributors.\label{fig:population-repo-teamsizes}]
      {
        \includegraphics[width=0.47\textwidth]{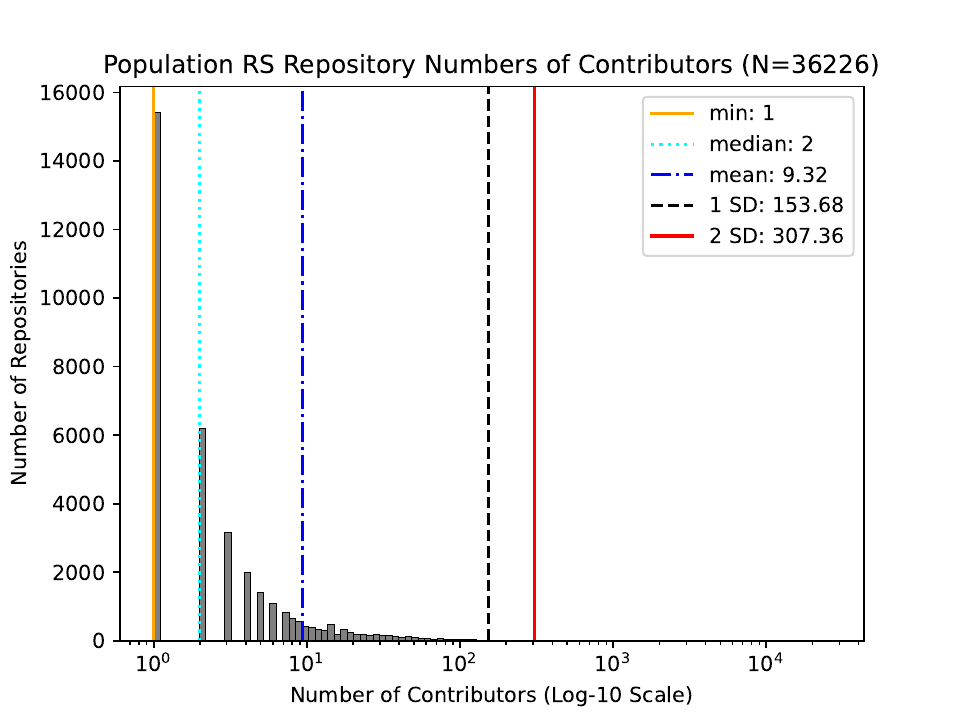}
      }
      \quad
      \subfloat[Distribution of sample repository log numbers of contributors, showing minimum of ten contributors (due to sampling exclusion criteria) and mean of around 100 contributors (96.04). \label{fig:sample-repo-teamsizes}]
      {
        \includegraphics[width=0.45\textwidth]{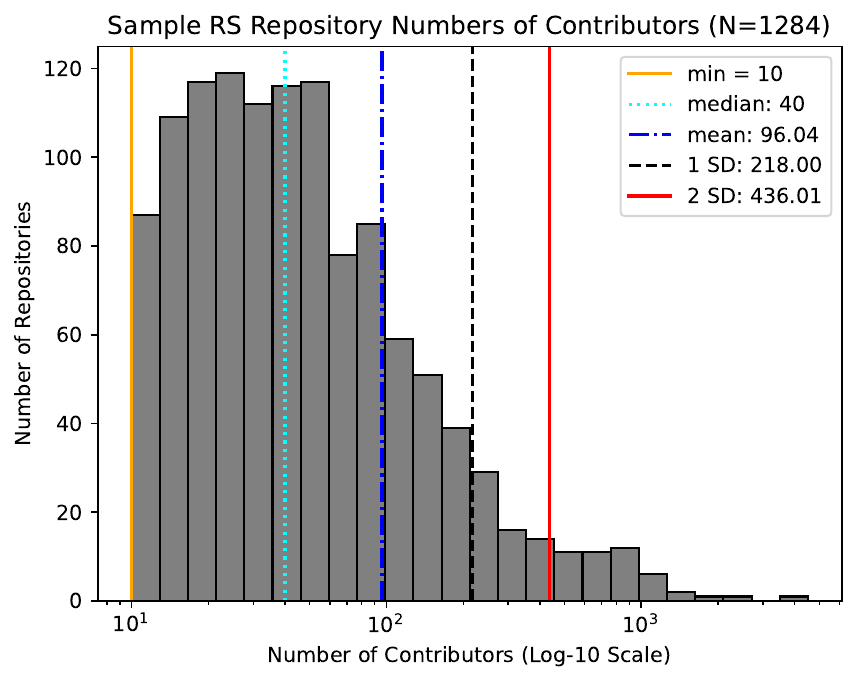}
      }
      \caption{Repository sizes (in numbers of contributors) at population and sample levels.}
    \end{figure}

    \subsubsection{Characterising the Study Sample Dataset}\label{TheSample}
    
    3,934,014 interactions were analysed in this study, made by 115,174 repo-individuals (96,947 unique GH usernames, with 10.40\% of usernames appearing in more than one sample repository). 
    These were made across the 1284 repositories (representing 3.54\% of the population set) and represent a substantial dataset for investigating the six included types of interaction (see Table~\ref{tab:interactions-sample-sums}).

    Figure~\ref{fig:sample-repo-teamsizes} shows the number of repositories containing different numbers of repo-individuals, and shows that the mean repository contained nearly 100 repo-individuals (mean=96.04) per repository. 
    Repositories with larger numbers of repo-individuals in the sample reached a maximum of 4,509.
    This was achieved despite the upper limit on number of unique committers to repositories in the sampling criteria because these criteria only limited the unique number of committers, and not repo-individuals who contributed through other types of included interaction, such as issue ticket creation (found to be the most frequent interaction type, seen in \autoref{fig:upset-sample}).

    Sample repositories show a mean of 131 commits made within the last year, while mean numbers of open and closed issue tickets were 56 and 555 respectively - a large increase compared to the wider population rates.
    This suggests greater interactivity within the sample than the overall population before sampling which is accounted for by the pooled effort and greater use of development management features in larger, collaborative teams.


    \paragraph{Interactions Patterns and RSE Personas from Clustering}\label{Clusters-Personas}
    
\begin{figure}
        \centering
        \subfloat[Initial clusters separate on high or low UIT, then high, medium or low MRC, to generate three sub-clusters.\label{fig:flowchart-intial-clusters}]{
        \includegraphics[width=0.55\textwidth]{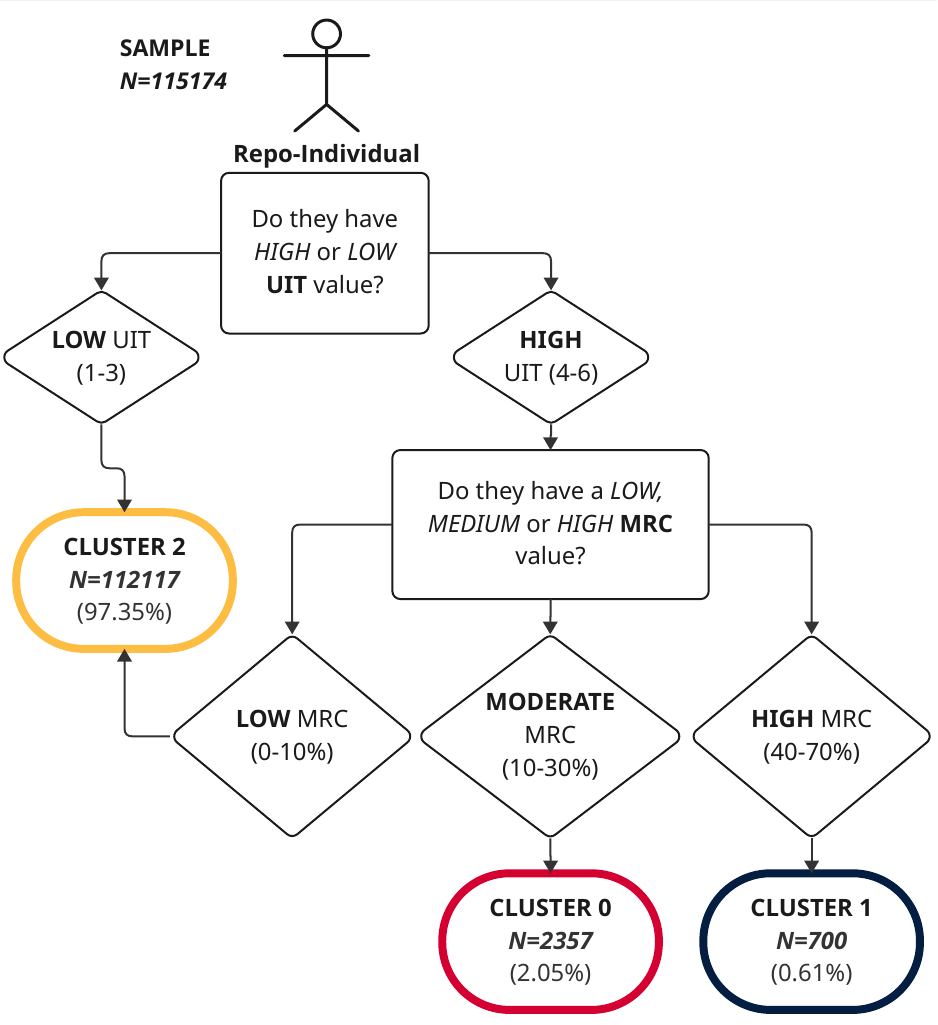}
        }
        \quad
        \subfloat[Moderate Interactivity Grouping (Cluster 0) contains two sub-clusters, differentiated by high or low volume of pull request closure. \label{fig:flowchart-cluster0}]{ 
        \includegraphics[width=0.39\textwidth]{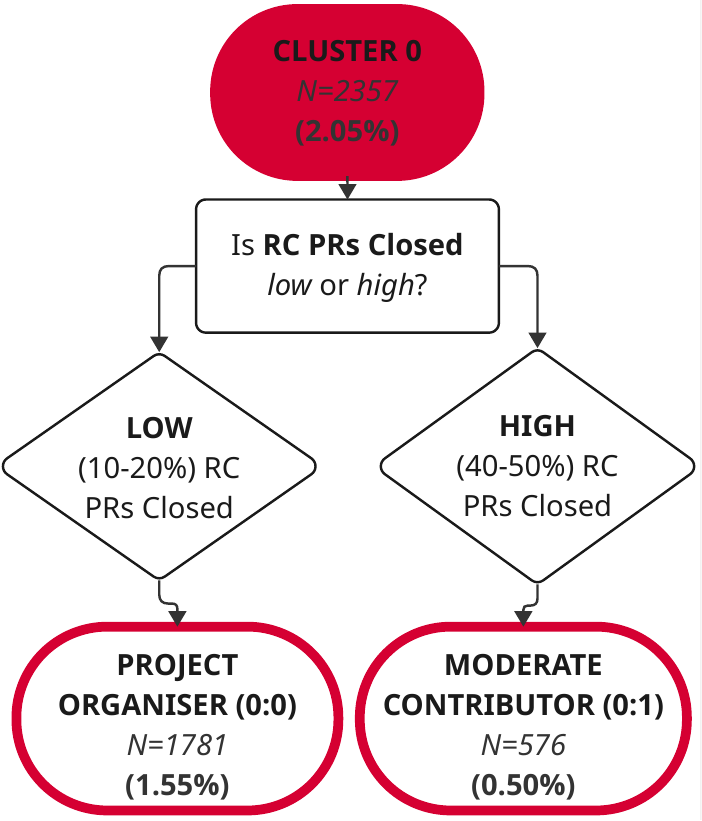}
        }
        \quad
        \subfloat[High Interactivity Grouping (Cluster 1): assignment frequency and commit creation determine three sub-clusters.\label{fig:flowchart-cluster1}]{
        \includegraphics[width=0.4\textwidth]{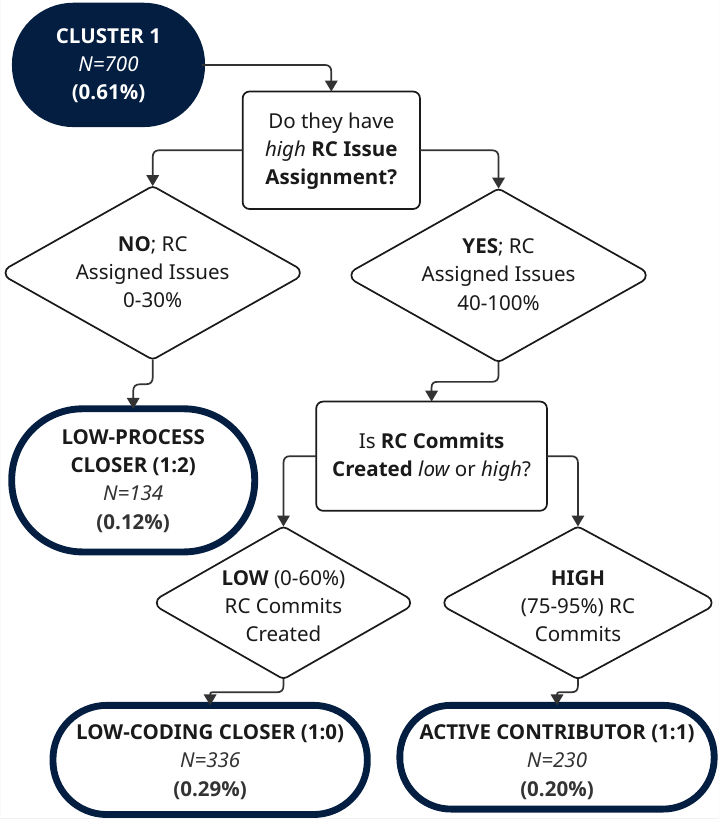}
        }
        \quad
        \subfloat[Low Interactivity Grouping (Cluster 2) shows either low or extremely low interactivity, via UIT, interaction days, and commit creation.\label{fig:flowchart-cluster2}]{
          \includegraphics[width=0.49\textwidth]{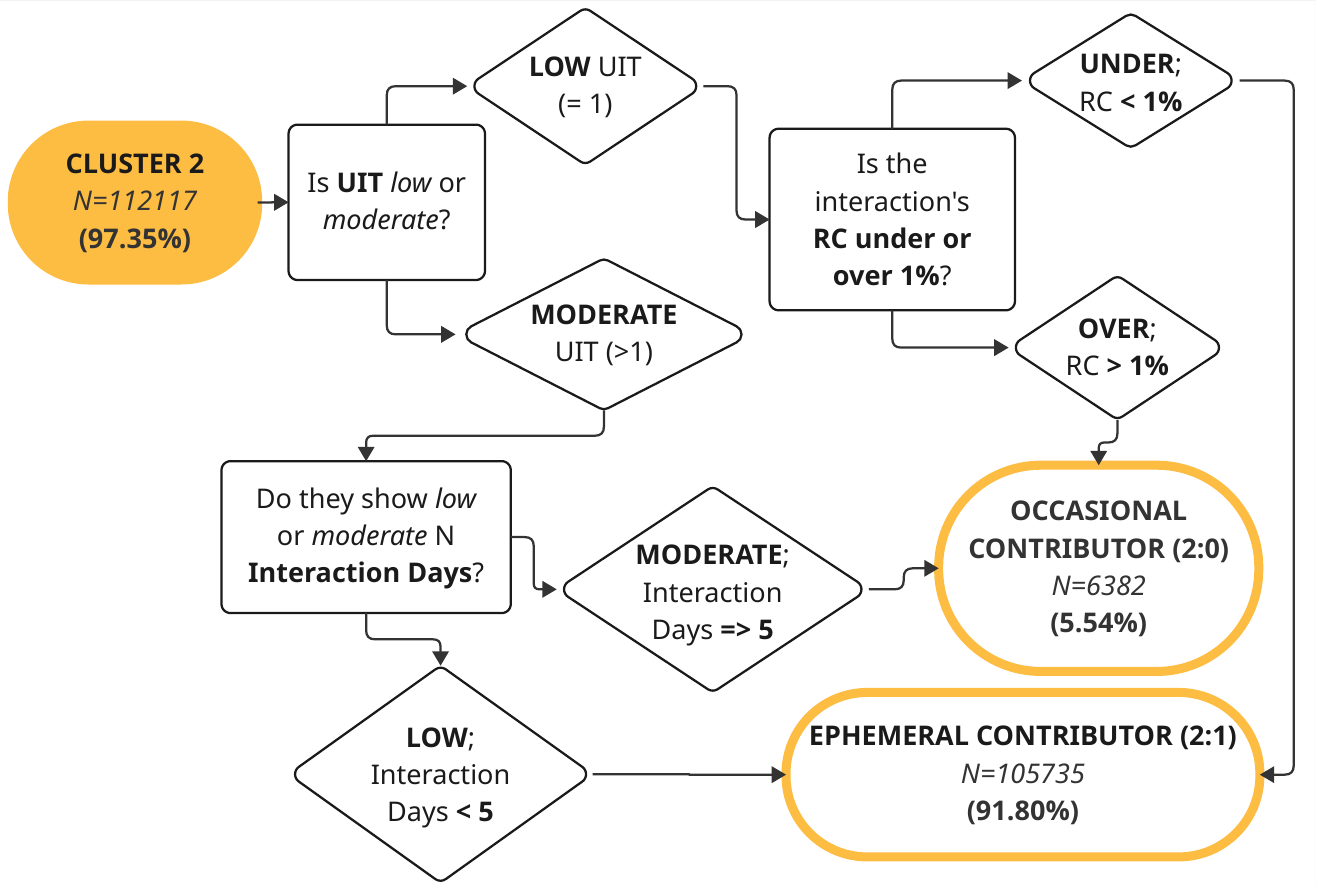}
          }
        \caption{Key variables splitting initial- and sub- clusters and describing RSE Persona differences, as well as proportions of sample occupied by each persona.}
    \end{figure}

    Hierarchical clustering analysis identified three initial clusters (Calinksi-Harabasz Index = 145,660.97 when N clusters = 3), as shown in Figure~\ref{fig:CH-index}.
    The relationships between these clusters can be seen visualised in the dendrogram in Figure~\ref{fig:dendro}, or as a flowchart identifying the key differentiating `splits' between these initial clusters in \autoref{fig:flowchart-intial-clusters}.
    
    As elaborated throughout these results, these initial clusters represent the same patterns of differences in both Variety of Interaction Behaviours (explored in Subsection~\ref{subsect: VarietyInteractions}) and Volume of Interaction Behaviours (reported in Subsection~\ref{subsect: VolumeInteractions}) by repository-individuals within the sample. 
    
\begin{figure}
    \centering
    \subfloat[Overview of Mean Repository Contributions (MRC) percentages per cluster: Cluster 1 show the highest contributions overall, while Cluster 2 demonstrate the lowest contributions averaged across all included types.\label{fig:MRC-counts-clusters}]
    {%
      \includegraphics[width=0.45\textwidth]{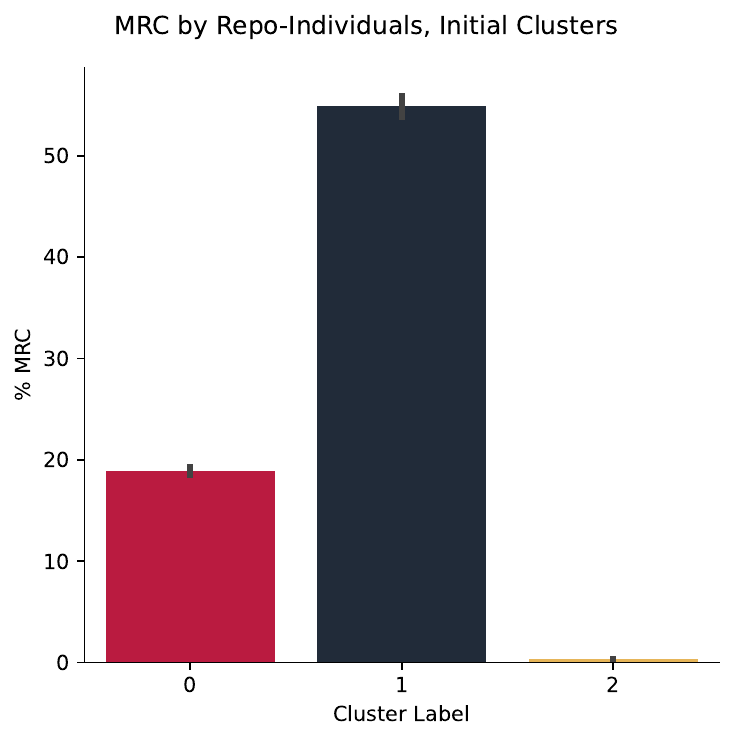}
    }\quad
    \subfloat[RC for Pull Request Closure: Repo-Individuals from Cluster 1 close almost 80\% of their repositories' pull requests. Cluster 2 contributors close less than 1\% of their repositories' pull requests.\label{fig:pc_PR_closed}]
    { %
      \includegraphics[width=0.45\textwidth]{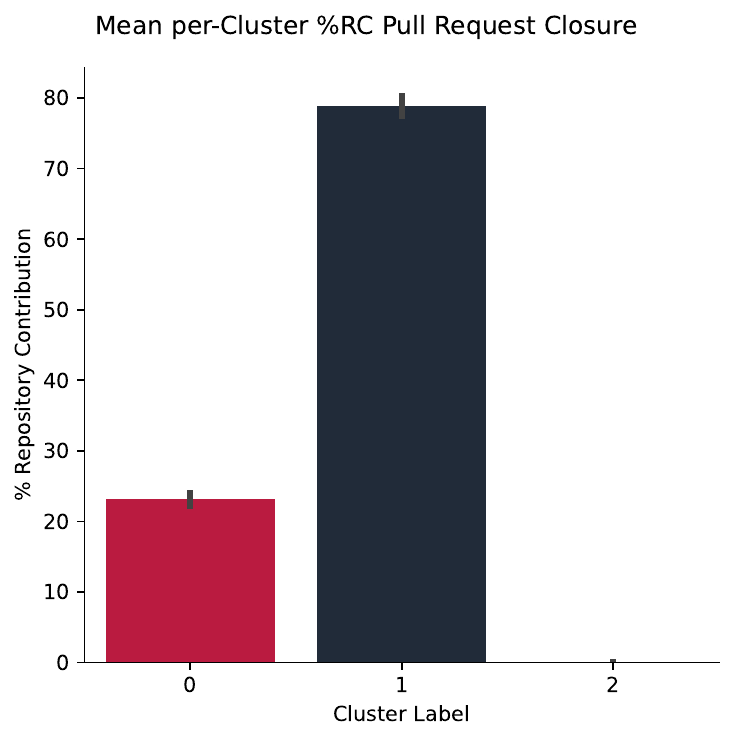}
    }
    \quad
    \subfloat[MRC values for all RSE Personas demonstrate mean normalised repository contribution percentages, equivalent to overall repository `impact' of a repository-individual from that persona.\label{fig:MRC-all-personas}]{
        \includegraphics[width=0.95\textwidth]{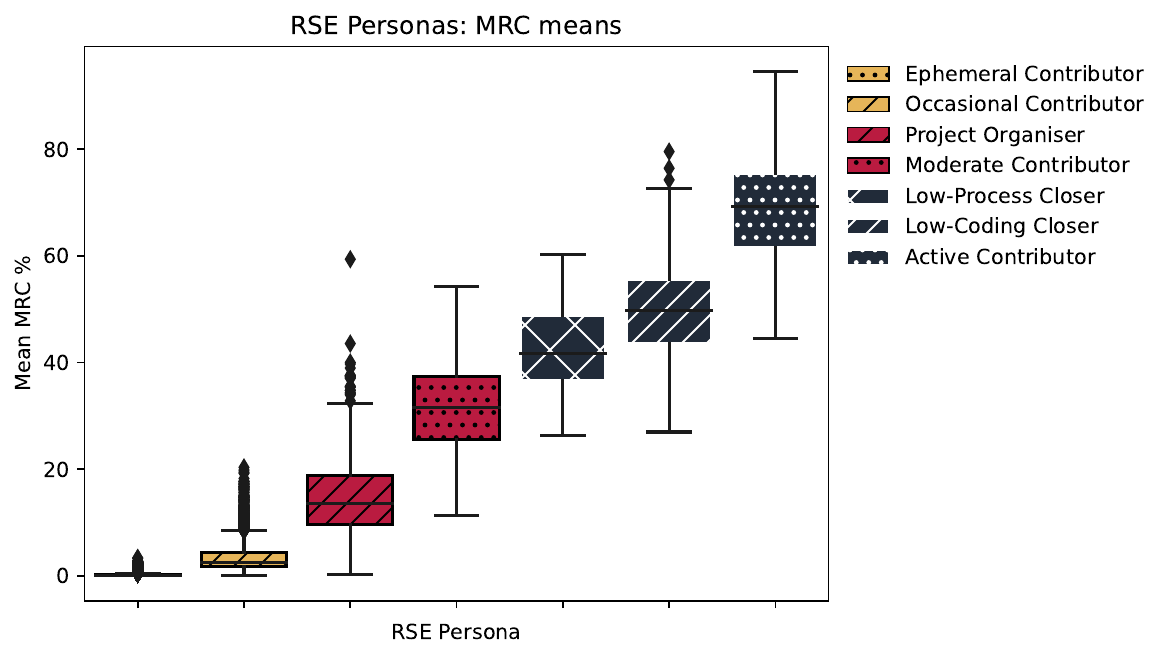}
    }
    \caption{Volume of interactions at initial cluster levels (\autoref{fig:MRC-counts-clusters} and \autoref{fig:pc_PR_closed}) and at sub-cluster levels (\autoref{fig:MRC-all-personas}) show the clear pattern of Low (Cluster 2), Moderate (Cluster 0) and High (Cluster 1) MRC values. This broad pattern is repeated from each RC variable (i.e. Pull Request Closure in \autoref{fig:pc_PR_closed}) and the differing volume of interactivity is a key separator for RSE Personas. \label{fig:volume-combo-figs}}
    \end{figure}

    For clarity, initial clusters are referred to as representing generally \textbf{High Interactivity (Cluster 1)}, \textbf{Moderate Interactivity (Cluster 0)}, and \textbf{Low Interactivity (Cluster 2)} contributors. 
    Figure~\ref{fig:UIT-counts} and Figure~\ref{fig:MRC-counts-clusters} respectively show overviews of differences between Variety of Interaction Behaviours as measured with UIT and Volume of Interaction Behaviours as described by MRC values.  

\begin{figure}[!tbp]
      \centering
      \subfloat[Low Interactivity Grouping (Cluster 2) represents over 97\% of all repo-individuals in the sample. \label{fig:cluster-distribution-repoindvs}]{
        \includegraphics[width=0.45\textwidth]{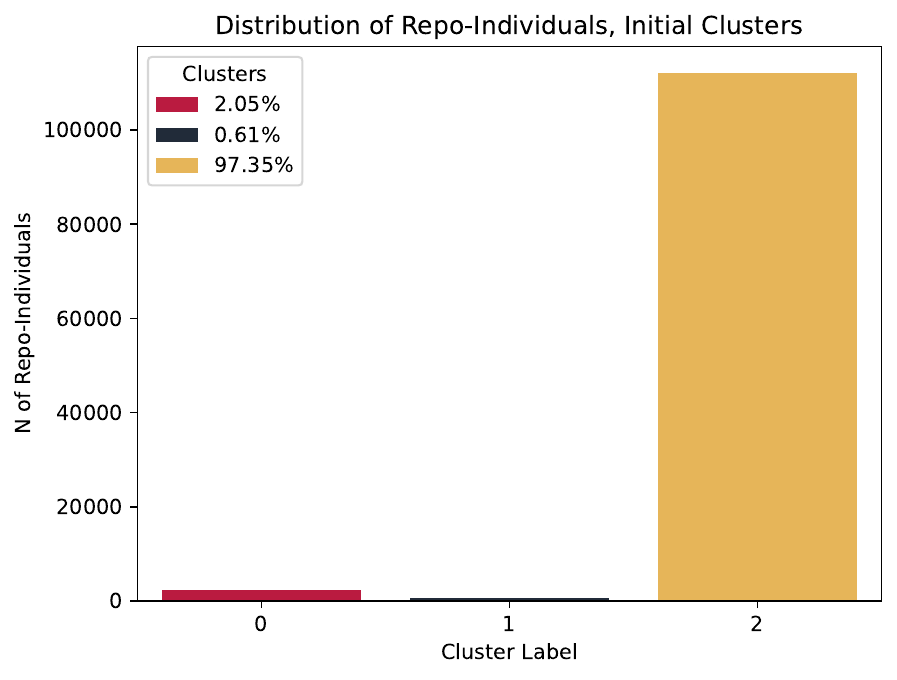}
      }
        \quad
        \subfloat[$\sim$ 60\% of repositories contain repo-individuals from at least two initial clusters; all three are represented in over a third.\label{fig:cluster-distribution-repos}]{
            \includegraphics[width=0.47\textwidth]{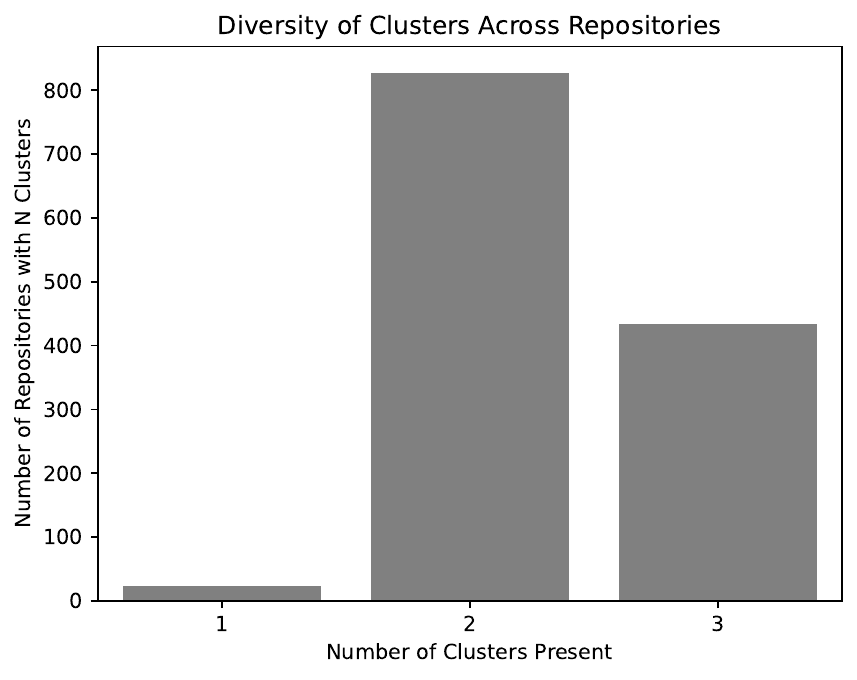}
        }
        \caption{Most repo-individuals are from low-interactivity groupings, while most repositories include individuals from at least two different interactivity groups.}
    \end{figure}

    Repo-individuals are not evenly distributed across these initial groupings, as can be seen in Figure~\ref{fig:cluster-distribution-repoindvs}.
    \textbf{The majority of repo-individuals (112,117; 97.35\% of the sample) were within the Low Interactivity Grouping} (Cluster 2). 
    Moderate Interactivity Grouping (Cluster 0) contained 2.05\% (2357) of the sample and High Interactivity Grouping (Cluster 1) contained 0.61\% of the sample (700). 
    Cluster 2 (Low Interactivity) also was also present in the majority of sample repositories (1276), followed by Cluster 0 (Moderate, in 1003 repositories), and Cluster 1 in 700 repositories - meaning that \textbf{each repo-individual assigned to Cluster 1 (Highest Interactivity personas) was in a separate repository}.
    Most repositories contain repo-individuals from at least two groupings, showing higher diversity of interaction-levels represented by the clusters (as seen in Figure~\ref{fig:cluster-distribution-repos}) suggesting `high-low' diverging interactivity amongst their team members. 

    \begin{figure}
        \centering
        \includegraphics[width=0.95\textwidth]{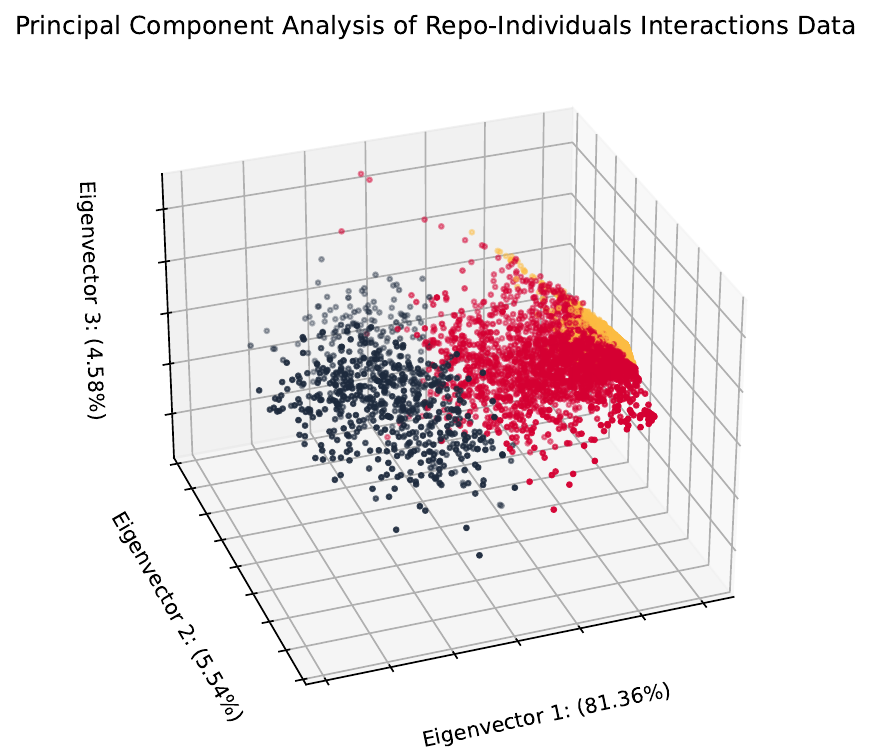}
        \caption{Principal Component Analysis results shows greatest separation on the first axis, explaining 81.36\% of variance. Subsequent axes only add 10.12\% of variance, bringing the total to 91.44\%. Low Interactivity Grouping (Cluster 2) (yellow) shows the tightest clustering, and significant overlap with Moderate Interactivity Grouping (Cluster 0) (red).}
        \label{fig:pca-3d}
    \end{figure}
   
    The high percentages of variance (86.90\% on the first two axes, and 91.48\% when including the third axis) explained by the Principal Component Analysis in Figure~\ref{fig:pca-3d} suggests the majority of factors which explain differences between the clusters have been captured in analysis, leaving only 8.58\% unaccounted for.
    
    While additional factors may assist in capturing and explaining further variance, those selected were effective enough to describe the majority of difference within the data. 
    However, the overlap seen between clusters 0 and 2 in Figure~\ref{fig:pca-3d} does indicate that these clusters may either require additional variables to separate them more clearly, or may show no distinct `real world' separation, with lots of travel by repo-individuals between these clusters.

    We can see this similarity between Clusters 0 and 2 (Moderate and Low Interactivity Groupings) in the dendrogram in Figure~\ref{fig:dendro} - both clusters derive from the same branching point with only 2,000-3,000 distance units between them and the horizontal branch they derive from, as opposed to Cluster 1 (High Interactivity), which has high vertical distance (around 5,500 units of distance) from the initial branching point (top line within the figure) and is therefore much more dissimilar. 
    
    Nevertheless, we can see from the higher CH Index score for N=3 clusters in Figure~\ref{fig:CH-index}, and the clear divergence on the dendrogram that there is still a distinguishable difference between Clusters 0 and 2, confirmed by the significant one-way ANOVA results for many variables between these initial clusters, and particularly amongst highly ranked features identified by the Feature Importance Analysis results in Table~\ref{tab:feature-importance}.   
    These results suggest that there is a robust basis for using the initial clustering results to describe RSE Personas.

    \begin{table}[]
    \centering
        \resizebox{0.6\textwidth}{!}{%
        \begin{tabular}{ |c | l | c | c |}
        \hline
            \multicolumn{1}{|l|}{\textbf{PCA}} & \multicolumn{1}{l|}{\textbf{Variable}} & \multicolumn{1}{l|}{\textbf{Importance Value}} & \multicolumn{1}{l|}{\textbf{Rank}} \\ \hline
            \textbf{PCA\_1}                    & \textbf{RC pull request closed}               & \textbf{39.81}                                       & \textbf{1}                         \\
            PCA\_1                    & RC issue closed                       & 37.50                                       & 2                         \\
            PCA\_1                    & RC sum N interactions                & 33.58                                       & 3                         \\
            \textbf{PCA\_2}                    & \textbf{RC commit created}                     & \textbf{27.74}                                       & \textbf{1}                        \\
            PCA\_2                    & RC sum N interactions                & 20.66                                       & 2                         \\
            PCA\_2                    & RC pull request closed               & 20.45                                       & 3                         \\
            \textbf{PCA\_3}                    & \textbf{RC issue closed}                       & \textbf{46.08}                                       & \textbf{1}                         \\
            PCA\_3                    & RC issue created                      & 37.45                                       & 2                         \\
            PCA\_3                    & RC pull request closed               & -1.90                                       & 3                         \\
        \hline
        \end{tabular}
        }
        \caption{Feature Importance Analysis ranking and percentage scores for top 3 (of 10) variables used in clustering across first three Eigenvectors.}
        \label{tab:feature-importance}
    \end{table}

    Each initial cluster was subset out and data re-clustered.
    Figures \ref{fig:flowchart-cluster0} to \ref{fig:flowchart-cluster2} demonstrate not only the proportions and sizes of different sub-clusters compared to the sample size. 
    Low Interactivity sub-clusters make up the majority of RSE Personas assigned to the sample repo-individuals for example, while all High Interactivity sub-clusters each only represent under 0.30\% of the sample
    These Figures also show features of the data important in separating the sub-clusters.

    \subsection{Variety of Interaction Behaviours}\label{subsect: VarietyInteractions}

    The mean number of unique interaction types (UITs) across all repo-individuals within the sample was 1.76 (indicating contribution to under two studied types), however, this increased when calculated as the mean of means by repo to 2.18.
    
    Types of interaction contributed by repo-individuals was shown to vary significantly across initial clustering (ANOVA p $<$ 0.05; Tukey's HSD p $<$ 0.05 for each combination of cluster mean UITs).

\begin{figure}[!tbp]
      \centering
      \subfloat[UIT counts per cluster show differing relationships: Cluster 0 (Moderate Interactivity) and 1 (High) show similar trends towards wider variety of interaction types, while Cluster 2 (Low Interactivity) shows a decrease in frequency at higher UIT values. Note: y axes show different scales.\label{fig:UIT-counts-clusters-detail}]{
      \includegraphics[width=0.95\textwidth]{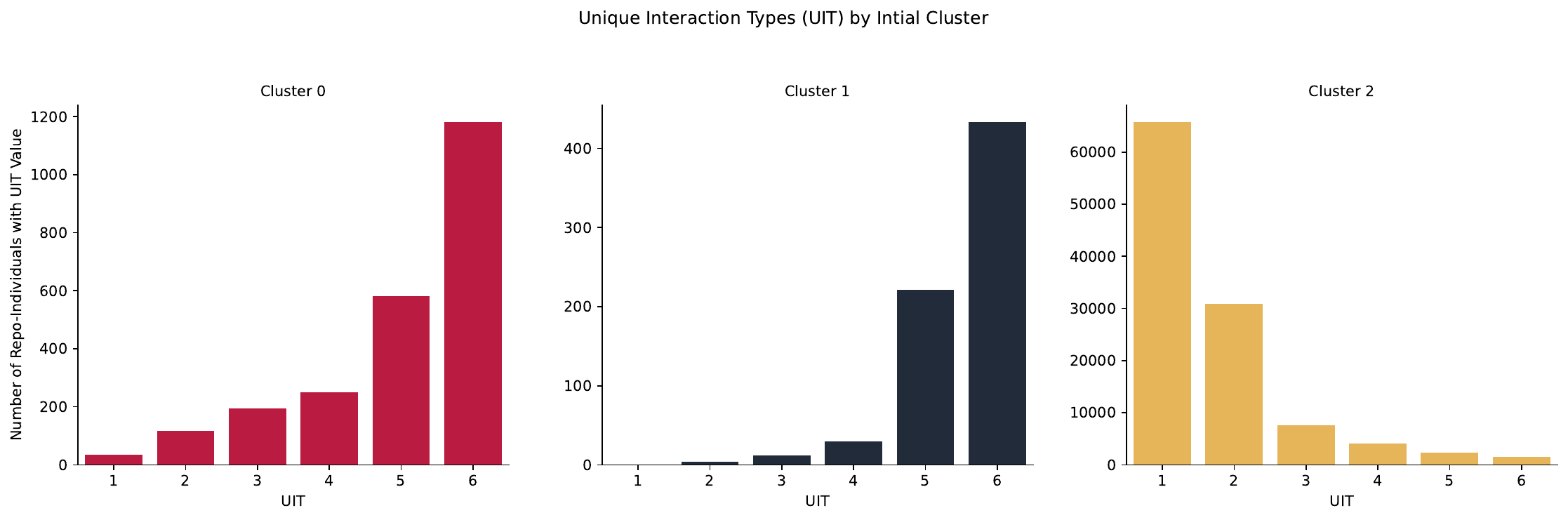}
      }
      \quad
      \subfloat[Sample repo-individual frequency at Unique Interaction Types (UITs) show most repo-individuals only contribute one interaction type.\label{fig:UIT-counts}]{
      \includegraphics[width=0.24\textwidth]{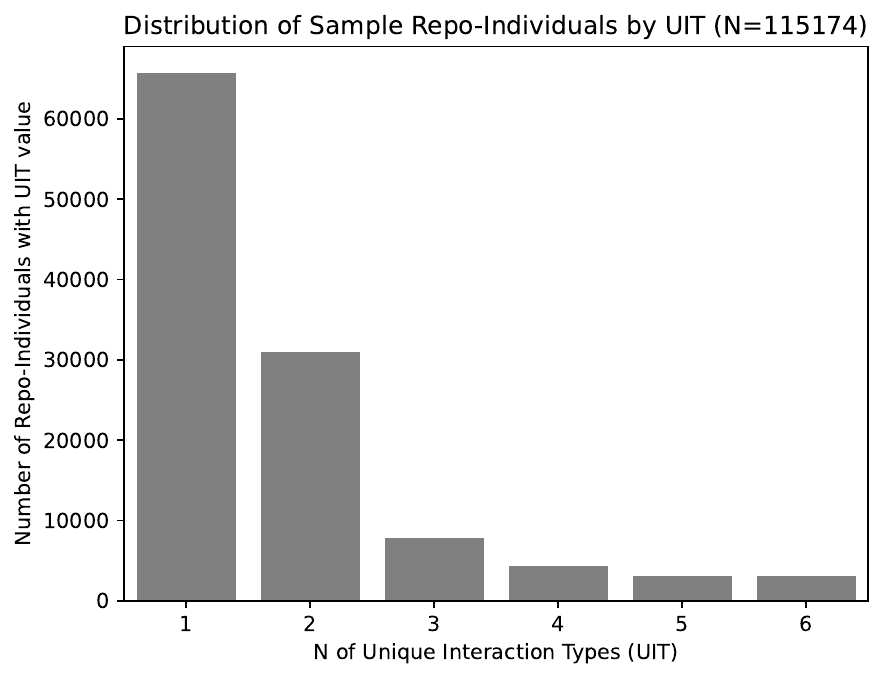}
      }
      \quad
      \subfloat[UpSet Plots show a matrix of combinations of interaction types and their frequency (\textit{upper bar-plot}), and overall numbers of repo-individuals contributing to each interaction type (\textit{left bar-plot}). Issue creation only is most common, while combinations including issue assignment are least frequent. Note: plot shows combinations made by $\ge$ 0.005\% of the sample. \label{fig:upset-sample}]{
      \includegraphics[width=0.7\textwidth]{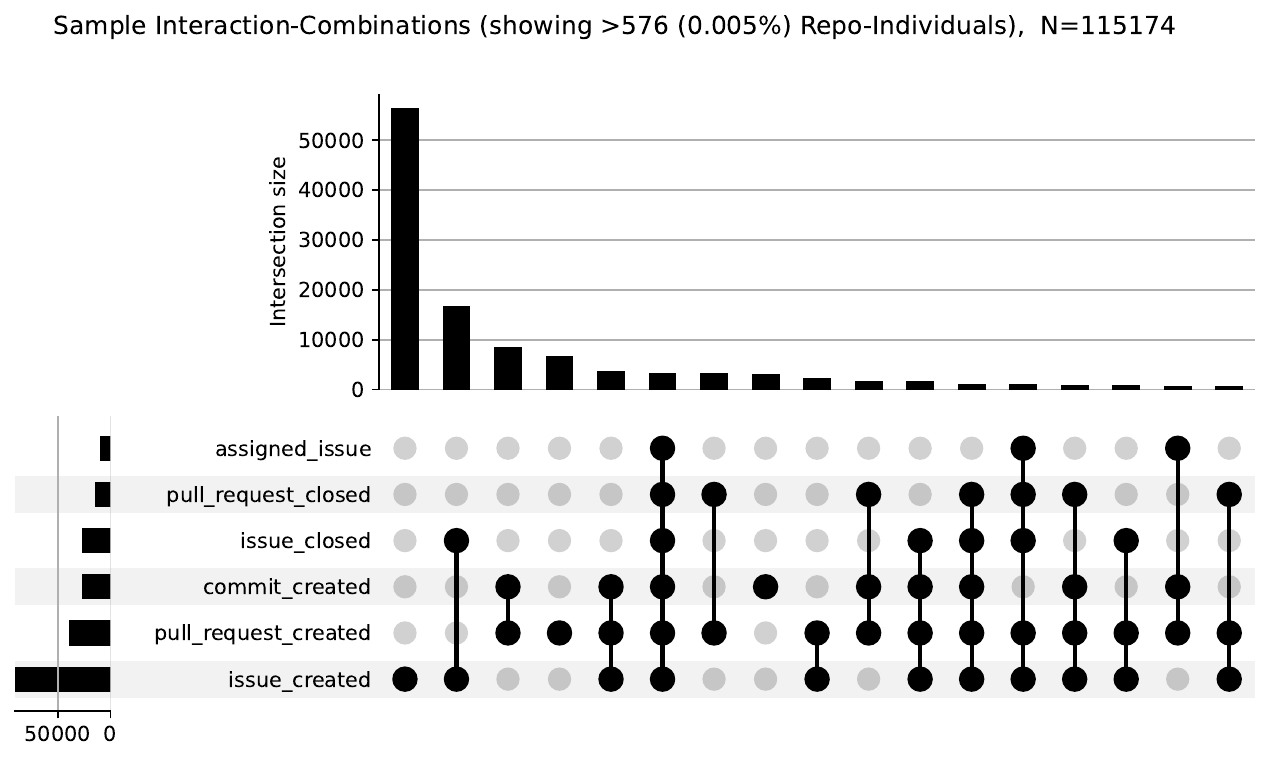}
      }
    \caption{Exploring variety of interactions: the majority of repo-individuals create Issue Tickets, and interactivity groupings Moderate and High demonstrate high variety of interactions, while Low Interactivity Cluster 2 shows narrow variety.}\label{fig:variety-combo}
    \end{figure}

    \textbf{Narrow interactivity (users who contributed to few types of interaction) is most frequent}, with the majority of repo-individuals generating only one or two types of interaction (see Figure~\ref{fig:UIT-counts-clusters-detail}).
    
    This is corroborated when looking at combinations of interaction types demonstrated within the dataset (Figure~\ref{fig:upset-sample}). 
    We also know that the majority of these repo-individuals are labelled as falling into Low Interactivity Grouping (Cluster 2). 
    This indicates narrow interactivity is commonplace in this group.

    Meanwhile, we see a trend from Figure~\ref{fig:UIT-counts-clusters-detail} within Clusters 0 and 1 where contribution to all or nearly all included interaction types is most common amongst these repo-individuals.
    ANOVA results confirm that the variety of interactions were significantly different (F: 16394.17, p: 0.0, p $<$ 0.05), while Tukey's HSD shows that each cluster significantly differs from each other cluster for UIT.

    Table~\ref{tab:interactions-sample-sums} also shows that interactions are not contributed equally across the overall sample:  \textbf{the majority of interactions are commits (55.22\% of all interactions)}, followed by pull request creations (11.99\%). 
    This may seem counter-intuitive if we recall that the most common interaction combination is `Issue Ticket Creation Only' (as seen in \autoref{fig:upset-sample}).
    However, while it is most common for someone contributing to only one type of issue to create issues, those who make any combination of interactions including commits are likely to create commits in far greater numbers than issue tickets, thus accounting for over two million commit creation interactions within the sample dataset, compared to only approximately 350,000 issue ticket creations.  

    \begin{table}[]
    \centering
        \resizebox{0.75\textwidth}{!}{%
        \begin{tabular}{ |c | c | c |}
        \hline
            \textbf{Interaction Type} & \textbf{Sample Interactions (N)} & \textbf{Sample Interactions (\%)} \\
            \hline
            Issues Assigned & 193,588 & 4.92 \\ 
            Issues Closed & 282,613 & 7.18 \\ 
            Issues Created & 352,913 & 8.97 \\
            Pull Requests Closed & 460,965 & 11.72 \\
            Pull Requests Created & 471,677 & 11.99 \\
            Commits Created & 2,172,258 & 55.22 \\
            \hline
            \textbf{ALL TYPES} & \textbf{3,934,014} & \textbf{100} \\ 
            \hline
        \end{tabular}
        }
        \caption{Total numbers of interactions for all types within the sample.}
        \label{tab:interactions-sample-sums}
    \end{table}

    \begin{table}[]
    \centering
        \resizebox{1\textwidth}{!}{%
        \begin{tabular}{ |c | cc | ccc | cc |}
        \hline
        \multicolumn{1}{|l|}{Variable \textbackslash Mean Values (\%)} & \multicolumn{1}{c|}{\textbf{Sub-cluster 0:0}} & \multicolumn{1}{c|}{\textbf{Sub-cluster 0:1}} & \multicolumn{1}{c|}{\textbf{Sub-cluster 1:0}} & \multicolumn{1}{c|}{\textbf{Sub-cluster 1:1}} & \multicolumn{1}{c|}{\textbf{Sub-cluster 1:2}} & \multicolumn{1}{c|}{\textbf{Sub-cluster 2:0}} & \multicolumn{1}{c|}{\textbf{Sub-cluster 2:1}} \\ \hline
        \multicolumn{1}{|l|}{}                          & Project Organiser                               & Moderate Contributor                          & Low-Coding Closer                             & Active Contributor                                 & Low-Process Closer                         & Occasional Contributor                        & Ephemeral Contributor                         \\ 
        RC Commit Creation                                             & 10.70                                         & 30.13                                         & 32.88                                         & 83.19                                         & 74.95                                         & 3.23                                          & 0.06                                          \\
        RC Issue Creation                                              & 12.35                                         & 22.47                                         & 40.47                                         & 51.72                                         & 16.12                                         & 3.61                                          & 0.38                                          \\
        RC Issue Closure                                               & 12.96                                         & 43.62                                         & 74.85                                         & 82.59                                         & 72.68                                         & 2.07                                          & 0.08                                          \\
        RC Assigned Issues                                             & 22.71                                         & 34.83                                         & 65.48                                         & 77.09                                         & 11.93                                         & 2.76                                          & 0.03                                          \\
        RC Pull Request Created                                        & 15.55                                         & 25.56                                         & 36.74                                         & 50.92                                         & 37.04                                         & 5.40                                          & 0.20                                          \\
        RC Pull Request Closed                                         & 16.11                                         & 44.81                                         & 71.64                                         & 86.24                                         & 84.13                                         & 1.94                                          & 0.03                                          \\
        \hline       
        \textbf{MRC}                                                   & \textbf{14.85}                                & \textbf{31.32}                                & \textbf{50.08}                                & \textbf{69.10}                                & \textbf{42.54}                                & \textbf{3.41}                                 & \textbf{0.15}                                 \\
        \hline                                      
        \end{tabular}%
        }
            \caption{Summary of RSE Personas with values for key interaction variables: differences in interaction patterns within these initial Moderate, High and Low Interactivity umbrella clusters clearly separate the sub-clusters.}
            \label{tab:sub-clusters-results}
    \end{table}

    \subsection{Volume of Interaction Behaviours}\label{subsect: VolumeInteractions}

    \textbf{Volume of interactivity by repo-individuals in different Interactivity Groupings can be summarised by looking at MRC (Mean Repository Contribution)}.

    Figure~\ref{fig:MRC-counts-clusters} illustrates the `low interactivity' trend typifying Cluster 2: this cluster showed the lowest mean MRC value (MRC values averaged across all repo-individuals within Cluster 2) of 0.34\%.
    Cluster 0 repo-individuals exhibited a moderate overall percentage contribution to their repositories of 18.88\%.
    Cluster 1 demonstrated a significant contribution to their repositories; the MRC value was 54.89\%, therefore showing strong tendency towards `high interactivity'. 
    
    One-way ANOVA verifies the significance of these results: reporting an F statistic of 342100.14 and P value of 0.0 (p $<$ 0.005) when testing MRC against clustering groups.
    Tukey's HSD returned p values all below 0.005 when testing the difference between means of each set of clusters, indicating that each Cluster's MRC percentages are significantly different from each other. 
    A similar trend can be observed in Figure~\ref{fig:pc_PR_closed} for RC Pull Requests Closed.
    
    Examining RSE Personas MRC values in \autoref{fig:MRC-all-personas} neatly demonstrates the differing patterns of overall repository impacts displayed by the personas amongst each other, as well as their inter-cluster differences. 

\begin{figure} 
    \centering
    \subfloat[Assignment of Issues within High Interactivity (Cluster 1) Sub-clusters. Sub-cluster 1:2 has significantly lower RC assignment quartile values and mean (11.93\%) than Sub-clusters 1:0 or 1:1 which have mean values of 65.45\% and 77.09\% respectively. \label{fig:subcluster1_assigned}]
    {%
      \includegraphics[width=0.45\textwidth]{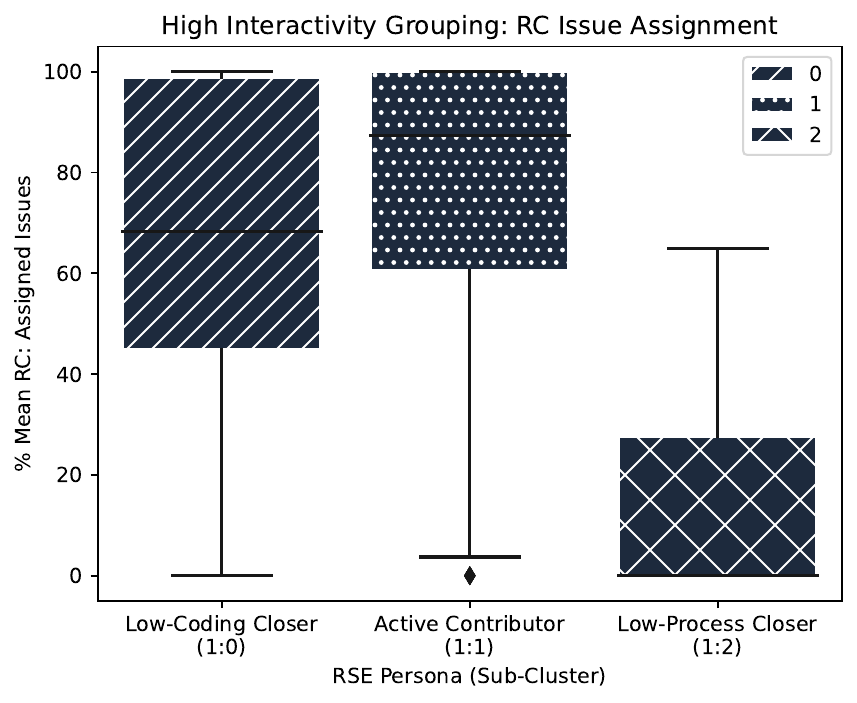}
    }\quad
    \subfloat[RC Commit Creation trends differ amongst Cluster 1 Sub-clusters: in this case, Sub-cluster 1:0 shows the lowest values, while Sub-clusters 1:1 and 1:2 are considerably higher. Mean RC Commit Creations is 32.88\% for 1:0, 83.19\% for 1:1, and 74.95\% for 1:2. \label{fig:subcluster1_commits}]
    { %
      \includegraphics[width=0.45\textwidth]{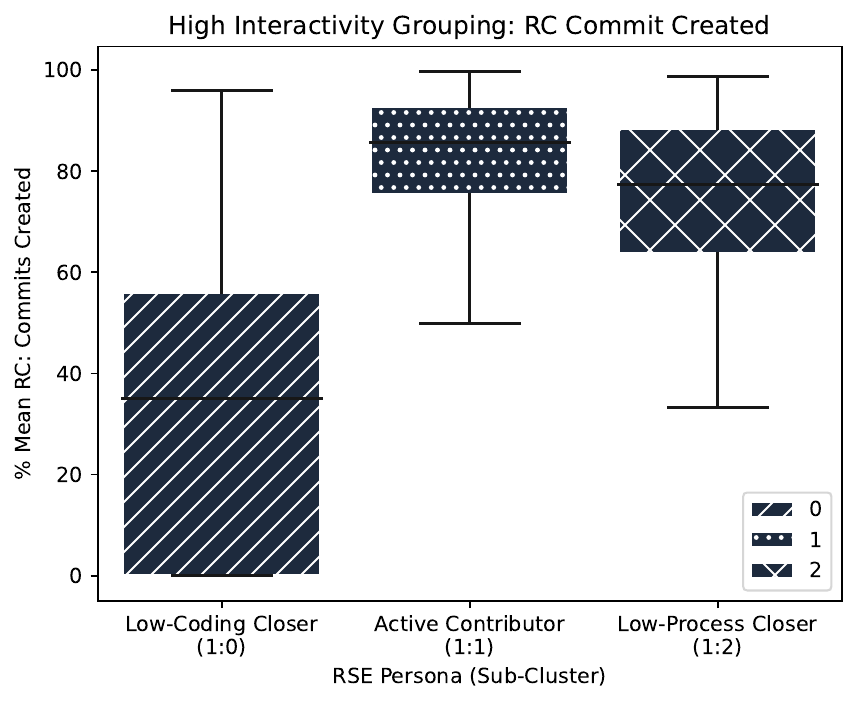}
    }
    \caption{High Interactivity Grouping (Cluster 1): key separating variables between sub-clusters are rates of Issue Ticket Assignment and Commit Creation.}\label{fig:high-group}
    \end{figure}

    \textbf{Issue ticket assignment is rare within the sample and represented the least frequent interaction type}, accounting for only 4.92\% of all interactions made, but as seen in \autoref{fig:flowchart-cluster1}, it is an important factor in differentiating between sub-clusters, and therefore between RSE Personas which may have similar interaction rates for other variables.
    To add further detail to the split seen in \autoref{fig:flowchart-cluster1}, \autoref{tab:sub-clusters-results} shows Low-Coding Closers and Low-Process Closers both demonstrate similarly high Issue Closure RC values of 74.85\% and 72.68\%, however their RC Assigned Issues are over 65\% and under 12\% respectively - indicating a large difference in their behaviour patterns for assignment which we can compare with Active Contributors in \autoref{fig:subcluster1_assigned}.

    Despite expectations of differentiated development activities amongst our clusters and potential personas, \textbf{neither commit classification method yielded significant results} when compared by cluster. 
    This strong negative result is surprising, given the strong differences in interaction types and volumes at a higher level, however, as data around commit classification was not included in the clustering, future work may find relationships.
    
    \begin{figure}  
    \centering
    \subfloat[Distribution of UIT values within Low Interactivity Grouping (Cluster 2) Sub-clusters, showing Sub-cluster 2:0 with higher UIT (a broader set of interaction types) than Sub-cluster 2:1.\label{fig:subcluster2_UIT}]
    {%
      \includegraphics[width=0.45\textwidth]{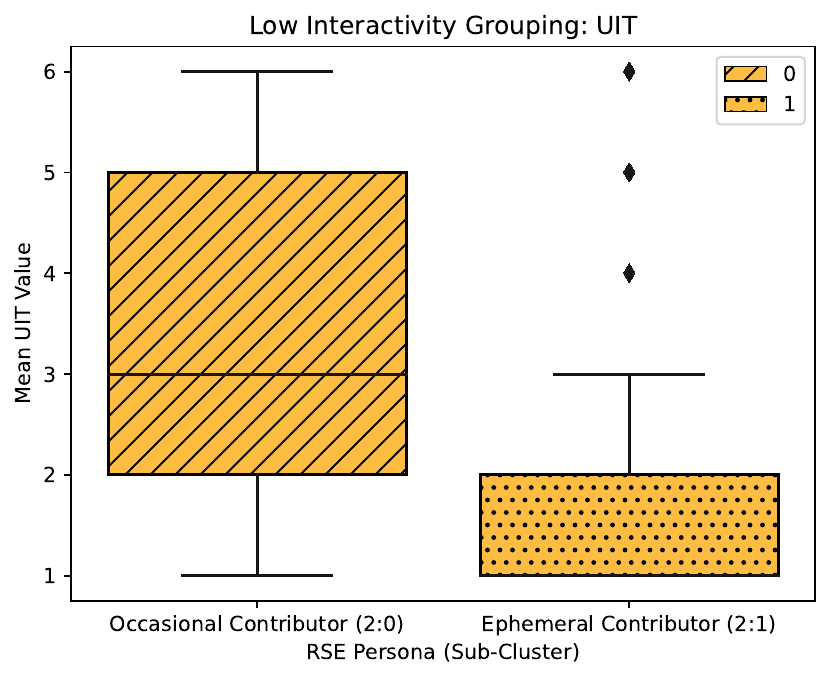}
    }\quad
    \subfloat[Overall MRC percentages for Cluster 2 sub-clusters show higher spread of outliers for Sub-cluster 2:0, while Sub-cluster 2:1 is tightly packed towards an overall contribution below 1\%.\label{fig:subcluster2_MRC}]
    { %
      \includegraphics[width=0.45\textwidth]{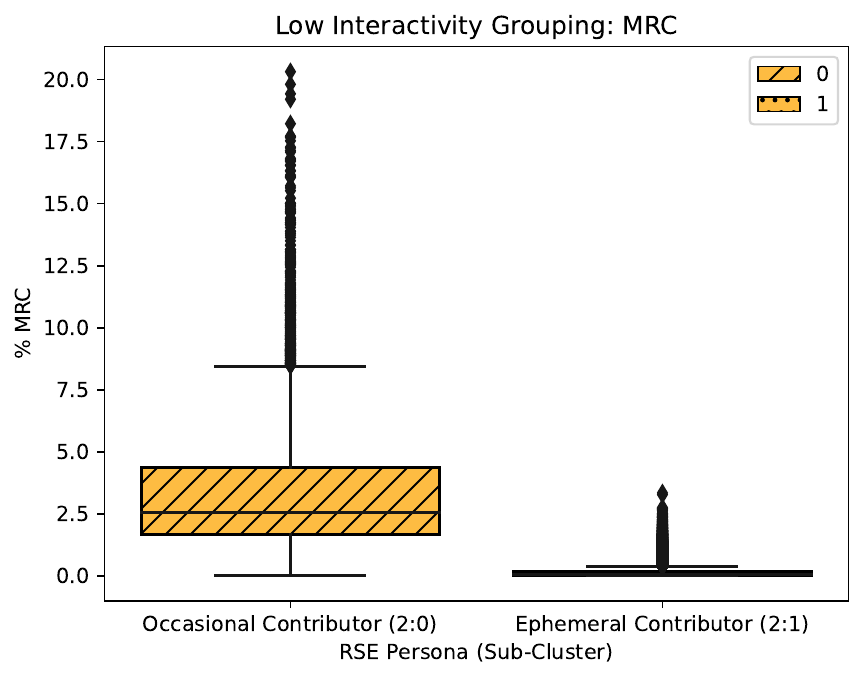}
    }
    \caption{Low Interactivity Grouping (Cluster 2) Sub-clusters show clear differences in their overall variety and volume of interactions.}\label{low-group}
    \end{figure}

    \begin{figure} 
    \centering
    \subfloat[RC Pull Request Closure demonstrates higher proportion of repository pull requests closed by Sub-cluster 0:1 than Sub-cluster 0:0 (mean 44.81\% versus 16.11\%). \label{fig:subcluster0_PRs-closed}]
    {%
      \includegraphics[width=0.45\textwidth]{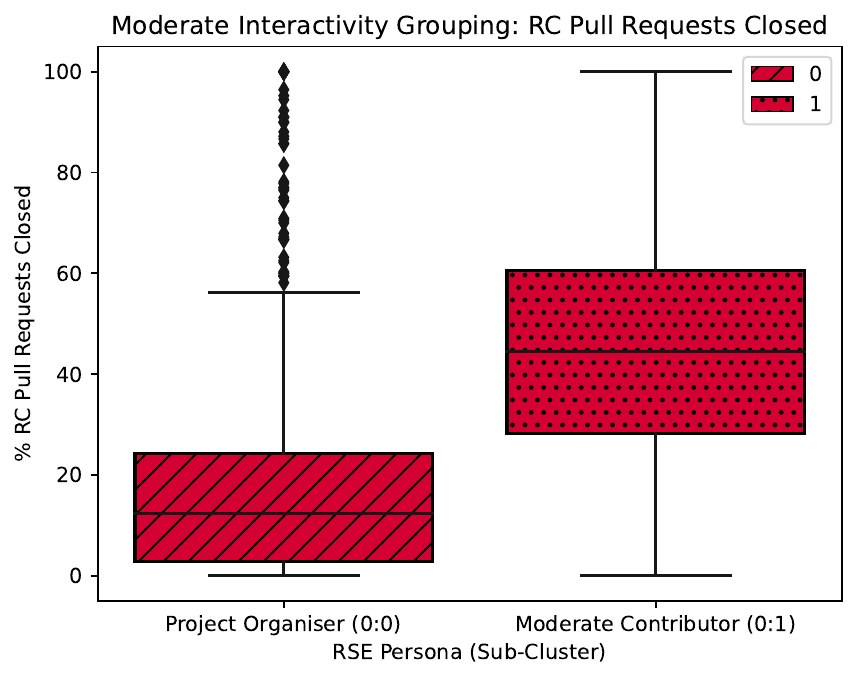}
    }\quad
    \subfloat[Following on from the higher rate of pull request closure, Sub-cluster 0:1 also demonstrates greater issue closure than opening, making them strong net issue closers (mean -43.40\% for 0:1 against -12.84\% for 0:0). \label{fig:subcluster0_created-closed}]
    { %
      \includegraphics[width=0.45\textwidth]{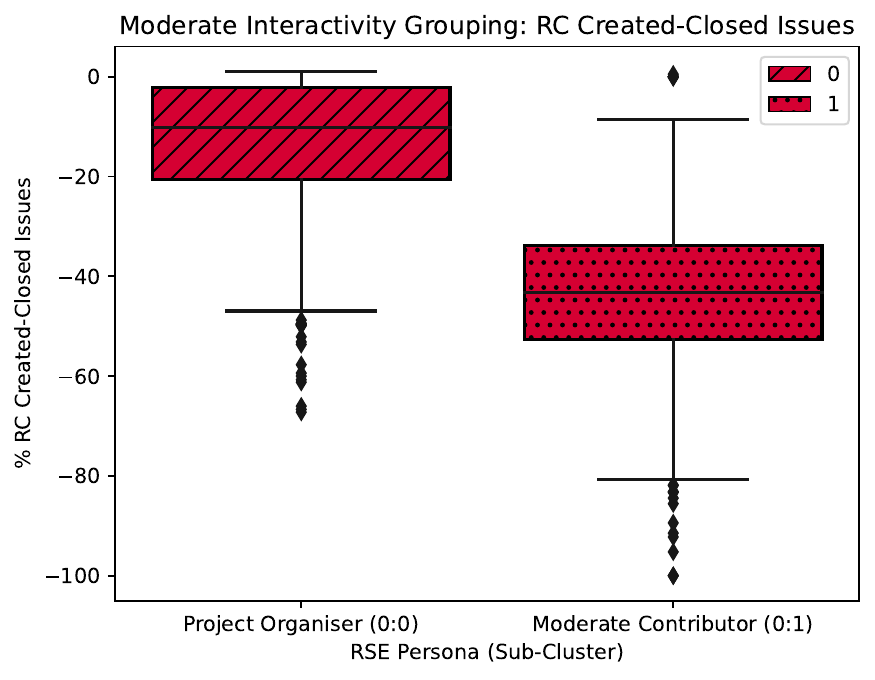}
    }
    \caption{Moderate Interactivity Grouping (Cluster 0) Sub-clusters are differentiated mainly on Pull Request and Issue Ticket and Net Closure Rates}\label{fig: mod-group}
    \end{figure}

    \subsection{Identified RSE Personas}\label{PersonasSummary}
  
    A summary of mean RC Interaction Type values shows the patterns of low and high interactivity in Table~\ref{tab:sub-clusters-results} and \autoref{fig:MRC-all-personas} for the sub-clusters, and demonstrates the separation between sub-clusters. 
    Ephemeral Contributor (0.15\%) and Occasional Contributor show very low values; Project Organisers and Moderate Contributors typify `Moderate Interactivity' personas; High Interactivity Personas `Low-Process Closer', `Low-Coding Closer' demonstrate around half of all interactions within their repositories, while `Active Contributors' show the highest MRC (69.10\%).
    
    High MRC values in \autoref{tab:sub-clusters-results} show that High Interactivity personas such as Active Contributors cannot co-exist within the same repository, as only one individual can contribute such high proportions of interactions within that repository. 

    While there is a clear trend of volume of interactivity separating the initial clusters in order, and MRCs differentiating the sub-clusters, there is also difference between sub-clusters amongst personas' most and least contributed interactions. 
    These distinguishing interaction types are described for each of the Interactivity Groupings (initial clusters) in Figures~\ref{fig:flowchart-cluster0} to \ref{fig:flowchart-cluster2} and help characterise their identified sub-clusters.
    
    For Sub-cluster 2:1 (Ephemeral Contributors), for example, we can see that the interaction types hypothesised as `high responsibility' and thought to be important in separating clusters are the lowest RC values: Assignment of Issues and Pull Request Closure (Figure~\ref{fig:flowchart-cluster2}). 
    There also seems to be some preference by low-interactivity personas towards Issue Creation interactions, as this type is found to have either the highest RC or second highest RC values for both Sub-clusters from Cluster 2 (Figure~\ref{fig:flowchart-cluster2}). 

    At the other end of the interactivity spectrum, for Sub-cluster 1:1 (Active Contributors, in \autoref{fig:flowchart-cluster1}) we see very high Pull Request Closure, and the highest RC figure for Assignment of issues of any persona.

    \subsubsection{RSE Persona: Ephemeral Contributor: Very Low Interactivity (2:1)}
    \textbf{Very narrow and shallow contributions made by visiting contributors.}
    This persona (sub-cluster 2:1) represents the overwhelming majority of the sample - 91.80\%  of the sample repo-individuals (105,735).
    This persona represents contributors who show extremely low UIT (mean 1.56, Figure~\ref{fig:subcluster2_UIT}) and low MRC (Figure~\ref{fig:subcluster2_MRC}) values, the lowest amongst all RSE Personas.
    Their single interaction type is typically issue ticket creation (55,869 repo-individuals), while the next most frequent combinations were `issues created and issues closed', followed by `commits created and pull request created'. 
    This persona shows almost no net issue creation or closure (mean RC created-closed issues is -0.08\%), and barely positive net pull request interaction (mean RC created-closed pull requests at 0.17\%), suggesting a very small overall impact. 
    They contribute only 0.23\% of mean RC Interaction Days in their repository, with an average of less than three interaction days suggesting their `ephemeral' (meaning `short-lived') title.
    This persona likely represents users visiting a GH repository to request a new feature or a fix. 
    Their frequency means that they are likely to contribute significant effort towards their repositories as a whole, and still play a valuable part within their RS projects despite their brief engagement.

    \subsubsection{RSE Persona: Occasional Contributor: Low Interactivity (2:0)}
    \textbf{Low contributions across a moderate variety of interaction types.}
    The smaller of the two low-interactivity clusters, representing 5.54\% of the sample repo-individuals (6382), this persona shows a mean MRC value of 3.41\%.
    This indicates a fairly low general contribution to their repositories by volume.
    Demonstrating wider engagement than Sub-cluster 2:1, it is not uniform - repo-individuals of this type appear to contribute Issue Ticket and Pull Request Creation far more frequently than Closure of these types.
    Mean UIT value is 3.42\% (Figure~\ref{fig:subcluster2_UIT}), showing somewhat low variety of interaction types. 
    The commonest UIT in this sub-cluster was found to be 2 (returned by 1650 repo-individuals, or 25.85\% of those with this persona), however the commonest combination of interaction types was found to be `all six types' (1001 repo-individuals, 15.68\%) - indicating the mean hides some bi-modality. 
    RC values are low in this sub-cluster, from 1.94\% for RC Pull Request Closed to a maximum of 5.40\% for RC Pull Request Created; the small range of 3.46\% amongst all interaction RC values suggests that only a few examples of contributions of each contributed types are made. 
    This persona demonstrates the `low UIT, low MRC' of the initially hypothesised `Occasional Contributor' persona identified in the pilot study, and keeps the name.
    This persona has weak mean net RC closure of issues (-2.03\%) and a moderate mean for RC Interaction Days of around 4\% of all interaction days towards their repository.
    This persona is likely to be an occasional contributor and a weak net opener of Pull Requests (mean RC created-closed pull requests is 3.45\%). 
    As with Sub-cluster 2:1, their relative frequency compared to rarer but higher-interaction personas means that they will be responsible for significant RS development within their projects.   

    \subsubsection{RSE Persona: Project Organiser: Low-Moderate Interactivity (0:0)}
    \textbf{Wide but shallow repository engagement: development managers rather than active developers.}
    1.55\% of sample repo-individuals (1781).
    This persona rates fairly low in most of the studied interaction types (Table~\ref{tab:sub-clusters-results}), contributing only around 10\% of their repositories' commits (an RC Commit Creation of 10.70\%).
    They seem to demonstrate higher assignment to issues: over 20\% of all assigned issues will go to repo-individuals with this persona. 
    As we define assignment to be a high-responsibility interaction type, this could indicate that these repo-individuals must be considered either capable of carrying out the tasks (despite relatively low Issue Closure and Pull Request Closure rates - the latter at 16.11\% in Figure~\ref{fig:subcluster0_PRs-closed}).
    Alternatively, it could mean this persona are assigned to tickets to keep them in the loop, perhaps because of some authority, importance or responsibility on the project overall. 
    The low rate of commit creation but high rate of high responsibility assignment of issues and low-to-moderate MRC values (mean 14.85\%), coupled with a mean UIT value of 4.88 out of 6, suggest that this persona could occupy the hypothesised persona of `Project Manager' which was expected at the outset of the RSE Personas pilot study (\cite{anderson_who_2025}).
    The Project Manager persona was expected to demonstrate a high UIT and low MRC: engaging in many types of interaction (e.g. usage of development management tools), but contributing relatively few, particularly code-related interactions, focussing on managing the project or coding effort rather than being centrally involved in development.
    Their net impact on issues is low overall closure (Figure~\ref{fig:subcluster0_created-closed}), despite the low coding volume and low MRC of 14.85\%.
    This sub-cluster represents a moderate volume of diverse interactions with focus on non-coding engagement, and as such are put forward as a persona with a strong project manager/organiser style, however we rename the initial hypothesised persona to `Project Organiser' to avoid confusion with the professional role of Project Manager. 
        
    \subsubsection{RSE Persona: Moderate Contributor: Moderate Interactivity (0:1)}
    \textbf{Moderately engaged, these contributors have varied interactions and good closure rates.}
    This persona contains 576 contributors, representing 0.50\% of the overall study sample. 
    There is a clear focus for this persona on getting closure: the highest RC values for this persona are Pull Request Closed (44.81\%, Figure~\ref{fig:subcluster0_PRs-closed}) and Issue Closed (43.62\%) -- nearly double the values for their equivalent Creation types (22.47\% and 25.56\%).
    As a result, these repo-individuals are good net closers: demonstrating RC Created-Closed Issues of -43.40\% (visible in Figure~\ref{fig:subcluster0_created-closed}), and RC Created-Closed Pull Requests value of -19.25\%.
    This persona type uses GH development management features frequently, and does the development work required to close repository tickets and pull requests, as demonstrated by their 30.13\% RC for Commit Creation.
    This contrasts with their sister persona of Project Organisers, who demonstrate similar UITs and breadth of interaction types, but show far lower commit volumes. 
 
    \subsubsection{RSE Persona: Low-Process Closer: Moderate-High Interactivity (1:2)}
    \textbf{Efficient fixers; closing existing PRs/Issues rather than engaging with development management processes to create new ones or `organise' efforts.}
    The smallest sub-cluster, represented by 134 repo-individuals: only 0.12\% of the study sample of over a hundred thousand contributors display this persona.
    These rare contributors show a mean MRC of 42.54\%, but this moderate value does not tell the full story.  
    This persona shows a fairly low RC value for Assignment of Issues and Issue Creation (11.93\%; 16.12\%), then the second highest mean RC of any persona type for Commit Creation and Pull Request Closure (74.95\%; 84.13\%) and third highest for Issue Closure (72.68\%). 
    The range between lowest and highest RC values is 72.20\%, showing a very strong behaviour preference away from Assignment (seen very clearly in Figure~\ref{fig:subcluster1_assigned}) and Issue Creation and towards Pull Request Closure.
    This interesting high-low pattern does not seem to be explained by considering high-responsibility interaction types, as we hypothesised that assignment to issue tickets and pull request closure were both high-responsibility types -- here we see low values for the former, high for the latter. 
    We suggest pull request closure is more closely linked to responsibility, as it is more likely to require fixing merge conflicts or checks to ensure the contributions addressing the pull request will benefit the project.
    Assignment to an issue typically refers to work `to be done' and is therefore less directly linked to changes than a pull request (typically work which has already been done).
    We see the highest rate of net pull request closure: the -47.10\% RC Created-Closed Pull Requests indicates that these repo-individuals are to thank for closing nearly half of all pull requests in their repository; the next highest persona achieves only -35.31\%.
    Net issue values are similarly strong, thanks to high closure and low opening rates for issue tickets: -72.52\%.
    The high Commit Creation RC seen in Figure~\ref{fig:subcluster1_commits} emphasises this `practical' approach to development.
    We see this repo-individual as more likely to get things done and close outstanding issues or pull requests, than creating new ones, and less involved in assignment of issue tickets.
    This persona represents fixers with less focus on using development management processes or tools for their own development tasks, hence the `low-process' moniker. 

    \subsubsection{RSE Persona: Low-Coding Closer: High Interactivity (1:0)}
    \textbf{High closure rates, but less coding: triages items and focuses on non-coding tasks.}
    This persona represents 0.29\% of the sample repo-individuals (336).
    They demonstrate strong RC values leading to a mean MRC of 50.08\%,and a mean UIT value of 5.43 out of 6, indicating high diversity of interaction types.
    More consistent in their RC scores than Sub-cluster 1:2, they demonstrating a range between highest and lowest values of only 41.97\%; a lowest RC value of 32.88\% (Commit Creation) and a highest of 74.85\% (for Issue Closure).
    The curiously low commit creation (in comparison to its' typically Highly Interactive Cluster 1 siblings, seen in Figure~\ref{fig:subcluster1_commits}) results are interesting: these repo-individuals show high rates of closure, but seem to achieve it with fewer commits than their colleagues in other high-interactivity personas. 
    There are several possible explanations for this unusual feature.
    Conciseness and brevity of coding style may contribute, as could potentially greater accuracy through use of linters and testing practices leading to fewer `typo fixes' or corrective commits. 
    But results of commit classification analyses do not provide evidence to support this: Hattori Lanza commit message classification did not identify any significant differences between sub-cluster means for corrective (fixing) commit activity, or any of the other development types examined. 
    Analysis of Hattori-Lanza commit size classification results found no significant differences, confirming that Low-Coding Closers don't solve their development tasks by using fewer, larger commits compared to other personas.
    Lower than expected commit rates with equivalent closure rates could point toward project management behaviours and may indicate this persona frequently `triage' issue tickets and pull requests: closing irrelevant or duplicated items, or completing items which do not need commit creation to resolve. 
    This is supported by a fairly high rate of Issue Ticket Assignment to these repository-individuals (Figure~\ref{fig:subcluster1_assigned}), potentially reflecting the 'informative assignment' pattern with low coding that we identified in the Project Organiser persona. 
    Overall, results indicate an infrequent but highly interactive and effective RSE persona with strong `closer' behaviour.    
    
    \subsubsection{RSE Persona: Active Contributor: Very High Interactivity (1:1)}
    \textbf{Core team members with the widest variety and deepest volume of contributions.}
    0.20\% (230) of sample repo-individuals are characterised by this persona derived from sub-cluster 1:1. 
    This persona demonstrates the most extreme high-intensity behaviour pattern, with the highest mean UIT value (5.88 out of a maximum of 6, with low standard deviation of 0.34). 
    Active Contributors show the highest overall mean repository contribution (MRC) in the dataset - 69.10\% of all studied interactions in these repo-individuals' repositories are made by them. 
    This persona therefore shows a clear grouping of highly engaged contributors, corresponding with the high MRC and high UIT properties hypothesised in the pilot study. 
    They demonstrate highest mean RC scores of all personas, across every included type (Pull Request Closed 86.24\%; Commit Created: 83.19\%, seen in Figure~\ref{fig:subcluster1_commits}; Issue Closed: 82.59\%; Assigned Issues: 77.09\%; Issue Created: 51.72\%; Pull Request Created: 50.92\%). 
    This persona shows the highest RC for Assignment of all sub-clusters, at 77.09\% of assignments made to these repo-individuals in their repositories, which is shown in Figure~\ref{fig:subcluster1_assigned}. 
    This persona also shows the highest Net Closure for Pull Requests with a value of -82.07\% for RC Created Minus Closed Issues.
    The mean RC interaction days these repo-individuals contribute towards their repository is significant: over 65\% of all interaction days are from this persona.
    Similarly, their contribution to their repository's total sum of interactions is nearly four out of every five contributions (a statistically significant 79.09\% for sub-cluster 1:1, compared to 54.10\% for sub-cluster 1:0 or 68.18\% for sub-cluster 1:2).
    These repo-individuals have the highest engagement values in the dataset and are undoubtedly important core members of their repositories. 
    They demonstrate consistently high usage of development management features on GH, impressive contributions towards the codebase through their commit creation (Figure~\ref{fig:subcluster1_commits}), and an active approach to RSE through high interactivity across the board.

\section{Discussion}\label{Discussions}

    This research attempted to prove that patterns exist within contributors' interactivity data for RS repositories on GH.
    Our results suggest that \textbf{not only do patterns exist amongst repository interactions, but that the RSE Personas method can successfully identify and characterise these patterns as distinct personas.}
    We have shown that three main levels of interactivity exist by contributors to the sample RS repositories: low, moderate and high interactivity.
    These can be shown in terms of types of interactions made by repo-individuals (described by significant differences in UIT across personas' members), as well as the scale of interactions of each studied type (summarised by MRC and shown in isolation by the significant differences between personas' RC percentage values). 
    We have additionally found that specific interaction types related to responsibility (Assignment to Issues, Pull Request Closure) are particularly important in separating sub-clusters to generate more detailed personas from broader initial clustering results made by focusing on the two former properties of interactivity (types and scale).
    
    Nevertheless, improvements to the RSE personas and further study will be valuable - in \autoref{subsect:limitations-threats} we discuss potential limits of our study, and outline areas where \nameref{FutureWork} can build on these. 

    \subsection{Limitations and Threats to Validity} \label{subsect:limitations-threats} 

    \paragraph{Variable Selection Choices}
    Supplementing studied interaction types may test whether identified personas remain robust, with the same repo-individuals clustering into the same personas after introducing additional interaction information such as code review, or issue ticket discussion interactions. 
    Alternatively, new types may disrupt current results, generating new personas with no relationship to those identified and described here. 

    In general, MRC acts as a reasonable indicator or summary for most personas described here.
    Nearly all personas showed fairly closely-grouped RC values for their interaction types, resulting in good description by the overall mean (MRC).
    Only in one persona (Low-Process Closer) did MRC seem to hide strong divergence in interactions behaviours; comparing MRC and standard deviations for each persona may be helpful in identifying this.
    It might also prove useful to analyse further subsets of the eligible sample data not selected for this study (due to computational limits), attempt to assign personas based on MRC-only scoring, then apply the full method and compare the accuracy of assignment. 
    If the results are accurate, using MRC as an indicator of RSE Personas could significantly decrease the time and computational effort required for the RSE personas analysis and improving its' potential for use, rather than applying the full data collection and analysis workflow applied by this research.

    We had expected commit classification methods to be effective in identifying differences in development behaviours between personas, however the lack of statistical significance between Hattori-Lanza commit message or size categories (\cite{hattori_nature_2008}) or Vasilescu et al. commit file-type categorisation (\cite{vasilescu_variation_2014}) suggests that if specialisms do exist between repo-individuals in the dataset, these do not necessarily vary relative to interaction patterns or RSE Personas.
    Including commit classification variables at the clustering stage in future could help further understand commit behaviours and whether they may still prove valuable to the RSE personas concept.

    The relationship between initial clusters identified by the dendrogram (Figure~\ref{fig:dendro} shows a closer relationship between Moderate and Low groupings, than Moderate and High) and the patterns identified by clustering and exploration of the RC and MRC values (where Moderate and High groupings seem closer, showing moderate or strong interactivity) do not on initial consideration seem to make sense.
    However, further investigation during development of the personas seems to indicate that scale of interaction is more important than initially suspected, and Cluster 1 represents Highly Interactive repo-individuals who are significantly atypical in their contribution intensity, compared to the more common low to fair interactivity and frequency of repo-individuals from Clusters 0 and 2.

    \paragraph{RS Projects Versus RS Repositories}
    The MSR methods applied provide many potential threats and perils as discussed in the comprehensive work by \cite{kalliamvakou_-depth_2016}. 
    One peril not addressed in this study relates to the definition of RS projects: this study evaluates RS at the repository level, discounting forks, while those authors advocate for working at the project level and suggest identifying all forks of the base repository and combining their data to avoid discounting valuable information in contributions made within forks. 
    Identifying forks of valid sample repositories and combining the data while avoiding duplication would be a significant undertaking, while we remain uncertain about whether interactions data from a non-merged repository could be validly compared with or collated with the base repository, as non-merged fork interactions may not be collaborative in the same way, reducing the accuracy of the dataset in addressing the aims of this research.

    \paragraph{Skew Towards Best Practice}
    This dataset represents a highly skewed subset of the RS population as a result of this study's sampling methods, however we acknowledge that selecting Zenodo as our source for potential RS repositories is likely to have selected in favour of RS projects already applying some level of best practices, as they have opted to deposit their code with a research repository and obtain a DOI. 
    This indicates that repository owners and contributors would have expected their code to be seen, which may also have prompted adoption of many collaborative `best practices' or usage of development management tools or features.
    The general nature of Zenodo for depositing research means that RS repositories obtained via Zenodo cannot be grouped into a particular academic field without further linkage and substantial work, and therefore represent a very wide potential range of research areas. 
    This could reduce applicability of results as we could encounter differing practices in RS development between academic domains which may diffuse the signals of patterns of behaviour.   
    Nevertheless, we consider the repositories and projects we have obtained via our methods valid and valuable for study, and consider that our contribution towards understanding interactions within such projects is valuable for research into RS, RSEs, and general SE research.

    \paragraph{Bots and RSE Personas}
    This research does not account for bots - while we could identify bots by name, we did not adjust the methods to exclude these. 
    This may skew results if their interaction behaviour is strongly different to their human colleagues, and further investigation should explore the distribution of the known bots across the identified personas.
    For example, the Low-Coding persona could have captured a high number of bots which automatically close issues or pull requests, leading to the high closure rates with low commit RC values.  
    
    \paragraph{Missing RSE Personas?} 
    The pilot study expected to identify a hypothesised `Focused Developer' persona type, which would have demonstrated low UIT values, but high MRC percentage. 
    This persona was not identified in the initial research, and was not expected - or identified -  here. 
    This may be an artifact of the methodology for calculating MRC: each RC value (a percentage) for all included interaction types are added, then this sum is divided by the number of interaction types studied (six).
    This means that a repo-individual who contributed 100\% of commits and 100\% of issues in their repository would find themselves with a relatively low MRC of 33.33\%. 
    Only by contributing consistently highly can a repo-individual obtain a very high MRC value, however this immediately excludes them from a low UIT score.
    While MRC has been demonstrated to work well in separating the sub-clusters within the dataset and is fairly effective in summarising interaction trends, the drawbacks outlined above mean that alternative analysis would be required to clearly rule out a `Focused Developer' persona with the described properties. 
    This implication should be addressed by future work, as quantifying `high development effort, low development-management feature usage' contributors is important.

    \paragraph{Not All Interactions Are Equal} 
    This study uses percentage Relative Contribution (RC) to included interaction types as a comparable measure between repo-individuals, quantifying their relative significance within their repository. 
    Calculating RC values provides some means of addressing that not all repository interactions occur with the same frequency (for example being assigned to an issue tickets is significantly rarer than merely creating one).
    Nevertheless, we have not applied any weighting to specific interaction types to counteract or represent rarity or importance.

    \paragraph{Quantity is Not The Same As Quality.} 
    This research does not engage with interaction quality, merely its' quantity. 
    This means that our study results cannot be applied to estimate effectiveness; we cannot correlate `many' contributions with `good' and `few' with `less-good'.
    The results of this study could, however, be combined in future work with other research approaches to explore whether certain RSE personas are associated with longer-lasting or more popular projects on GH, or are frequently found in `healthy' community projects.

    \section{Future Work}\label{FutureWork} 

    We discuss some potential areas of work which could build on this research, highlighting gaps and opportunities to link RSE Personas into wider study of RSE at a range of levels. 
    
    \paragraph{Improving and Evaluating RSE Personas}
    Replicating this study with different RS repositories on GH, or applying this technique to GitLab repositories could provide comparisons between the RS projects and RSEs found on differing collaborative development platforms.
    While our data-driven methods addresses our research questions well and explores the patterns of interaction behaviours we have aimed to study, we do not use a mixed-methods approach.
    Taking that approach in future work might identify the reasons for specific differences between the currently identified personas; for example, an interview approach to support the data-focussed aims of this study would be extremely helpful in addressing unexpected results such as Low-Coding Closers' low commit rates or Low-Process Closers' assignment avoidance.
    Collecting data on the `why' behind the `what' being explored in this research would establish a holistic basis for well-defined and useful RSE Personas.
    Assessing the extent to which repo-individuals did or did not identify with their labelled RSE personas through qualitative methods would strengthen the RSE personas concept significantly, and could provide feedback on which characterisations were found most or least useful by RSEs themselves which could feed into further development of the method and its uses.
        
    \paragraph{Hidden Contributors} 
    The data-driven nature of the RSE persona method means that some valuable contributions to the RS project are unlikely to be included if they do not overlap with included interaction types. 
    Currently, this includes discussions made on issue tickets by GH users who have not made any other interaction types - for example, this might represent researchers who have contributed theoretical or statistical elements upon which the code is based, or project managers who do not manage issue tickets by opening or closing them. 
    Inclusion of issue ticket discussion data as an additional interaction type would allow engagement by non-coding RS project team members to be captured and analysed - for example, researchers or project managers who may contribute domain knowledge, ideas, requirements, development management suggestions, or other information relevant to the project.
    These elements are not currently captured but these could be easily added to the dataset and combined with existing data to avoid excluding these important contributors.
       
    \paragraph{Comparing Apples with Oranges: Research Software Project Differences} 
    RS projects can take many forms, and researchers are currently exploring how RS can be categorised (e.g. \cite{hasselbring_categorizing_2024}).
    This study's method does not filter for RS repositories of one RS type (e.g. only infrastructure projects or data analysis workflows), nor do we select from one research domain only, resulting in a considerably varied sample set of repositories. 
    These were not manually analysed for software type or research domain as this would have involved considerable time and difficulty - particularly in the case of interdisciplinary research or general-purpose tools for research.
    Future work may consider using a RS categorisation approach in combination with RSE personas to explore whether different RS projects attract (or create) different development behaviours, and hence, RSE personas.
    
    \paragraph{RSE Personas Dynamics} 
    Developing a taxonomy and accompanying ontology for RSE personas would formalise the RSE persona concepts, explain relationships between personas with specific terminology and describe the properties of each persona in a reusable format. 
    This research did not dig into the distribution of personas for this publication, and this is a high priority for future work.
    The high percentages shown by personas within the High Interactivity Grouping (70\% to mid-80\% RC values) made it very unlikely that more than one high interactivity persona can exist within a repository, proven by our data (seen in \autoref{fig:cluster-distribution-repoindvs}).
    This poses questions around whether one is born, or `becomes' a persona: investigating persona stability over time by analysing repo-individuals' contributions over time may provide a way of exploring this.
    Likewise, testing whether repo-individuals commonly shift between personas would have valuable project management implications: if managers know how to support contributors in moderate or low interactivity personas to grow and develop into higher-interactivity personas after a rare high-interactivity RSE leaves a project, this could lead to more sustainable RS projects. 
    Addressing these points, and identifying which personas commonly co-occur will be valuable for exploring RS team dynamics and could establish the `community ecology' of RSE personas.
    Examining whether repo-individuals show the same or similar personas in other sample repositories they interact with would be another valuable analysis - enabled by our use of repo-individuals as the unit of study, rather than GH usernames. 

\section{Conclusion}\label{Conclusion} 

    \textbf{This paper describes the successful application of the novel, data-driven RSE personas method} to a large sample of RS repositories on GH and advances our understanding of common and rare contribution patterns by RSEs.
    \textbf{Seven RSE personas were identified} from 115,174 repo-individuals contributing to 1,284 RS repositories on GH to successfully address our research questions.
    
    We have explained persona relationships from initial clustering which identified \textbf{three levels of interactivity (low, moderate and high)}, and made comparisons with nearest sub-clustering neighbours.
    RSE personas describe distinct differences in both types (variety) and scale (volume) of studied interactions within their broad interactivity level.
    Use of \textbf{sub-clustering within majority and minority clusters successfully allowed us to further separate our initial clusters}, confirming this hypothesis (H3).
    We also proved that \textbf{pull request closure was a key interaction type which strongly influences RSE persona creation and differentiation (H2)}. 
    This study confirmed the existence of hypothesised `Active Leader' (renamed \textbf{`Active Contributor'}) and \textbf{`Occasional Contributor'} RSE personas from the pilot study results (H1). 
    Additionally, we confirm the hypothesised `Project Manager' persona which had not been found in the smaller pilot study, however we have renamed it \textbf{`Project Organiser'} to differentiate from the professional project manager role. 
    We also identified four new RSE personas: \textbf{`Ephemeral Contributor'}; \textbf{`Moderate Contributor'}; \textbf{`Low-Process Closer'}; and \textbf{`Low-Coding Closer'}.

    We hope to build on this method in future work by improving and evaluating the RSE personas method with complimentary research approaches and expanding the interaction types studied to identify potentially hidden contributors.
    We also aim to combine this method with RS or RS project classifications to determine to what extent project factors influence RSE personas, and to investigate persona dynamics more deeply to explore how RSE personas co-occur in RS project teams and what project management implications this may have.

    \textbf{We believe that this research builds a strong foundation for work on improving both individuals' and collaborative teams' RS development}, by characterising development and interaction patterns into RSE Personas in a way that RSEs and RS team managers can relate to and use  -- in conversations about skills, engagement and roles -- while gaining an overview of current patterns in collaborative RS development across the field.

\begin{acknowledge}
    We are grateful to the reviewers of this manuscript for their helpful comments, time and obvious care in reviewing this study. 
    We warmly wish to acknowledge the attendees of deRSE25 conference (Karlsruhe, Germany; 25-27th February 2025) for their enthusiastic and thoughtful engagement with the earlier poster presentation of this research. 
    EPCC colleagues also provided further helpful discussion on subsequent presentation of the same poster.
    Dr. Mark Bull (EPCC) assisted in developing the form of the project and ideas on methodology.
    The lead author acknowledges `Rubber Duck Research Club' for their mutual support, providing a sounding board for developing early ideas, and their patience during bug fixing sessions for the data collection and analysis code. 
    Thanks for minor manuscript edits for readability go to Tom Hermanovsky.
    This work was supported by a Doctoral Training Partnership award for project number 2662705, Coding Smart in Academia: Evidence-Based Software Engineering Approaches for High-Quality Research Software Projects, through grant EP/T517884/1. 

\end{acknowledge}

\bibliographystyle{eceasst}
\bibliography{references}

\section{Appendix: Code and Data Availability}\label{code-data}

    \paragraph{Dataset Availability}
    The data supporting this research is available at Zenodo: 
    \\ \texttt{10.5281/zenodo.15458472}.   
    
    \paragraph{Code Availability / Reproducibility Package} 
    The code used to collect, process and analyse the data, and generate many of the figures for this paper are available at the public GH repository for this project: \texttt{https://github.com/FlicAnderson/RSE-personas}, and the release used was \texttt{v1.0.2}.

    \subsubsection{Data Collection: Computational Environment}\label{subsubsect:comp-env}  
    A Virtual Machine (VM) was used to ensure a consistent and reproducible development environment for the data collection, processing and analysis, with 8CPU and 112GB RAM and 500GB storage access for data.
    This ran on Ubuntu Linux 22.04 LTS (long term support), with \texttt{conda} used for software package management.
    Reproducibility of this environment and version information of required software libraries and other software dependencies was ensured by using conda's environment file export and pip requirements file creation features (details can be found at the study GH repository). 
    Software developed for data collection, processing and analysis was written in \texttt{Python 3.10.9}, with PEP8 style guide followed, using \texttt{pylint} linting tool applied automatically in a VS Code IDE.
    Some testing was done locally using \texttt{pytest}.
    An additional GitHub Actions environment creation and linting- and test- running workflow using \texttt{pylint} and \texttt{pytest} was set up to run on code commit pushes to GH. 

\end{document}